\documentclass[aps,prx,twocolumn,colorlinks=true,linkcolor=black,urlcolor=blue,citecolor=black,pdfencoding=auto,longbibliography,superscriptaddress]{revtex4-1}
\usepackage{graphicx}
\usepackage{color}
\usepackage{amsmath}
\usepackage{amsfonts}
\usepackage{bbold}
\usepackage{amssymb}
\usepackage{tikz}
\usepackage{pgfplots}
\usepackage[colorlinks=true,
            linkcolor=black,
            urlcolor=blue,
            citecolor=black,
            pdfencoding=auto]{hyperref}
\usepackage{blindtext}

\newcommand{\rev}[1]{{\color{black} #1}} 

\begin{document}

\title{
Optical Spectral Weight, Phase Stiffness and $T_c$ Bounds for Trivial and Topological Flat Band
Superconductors
}

\author{Nishchhal Verma}
\affiliation{Department of Physics, The Ohio State University, Columbus, Ohio 43210, USA}
\author{Tamaghna Hazra}
\affiliation{Department of Physics, The Ohio State University, Columbus, Ohio 43210, USA}
\affiliation{Department  of  Physics  and  Astronomy, Rutgers  University,  Piscataway,  NJ  08854, USA}
\author{Mohit Randeria}
\affiliation{Department of Physics, The Ohio State University, Columbus, Ohio 43210, USA}

\begin{abstract}
We present exact results that give insight into how interactions lead to transport and superconductivity
in a flat band where the electrons have no kinetic energy.
We obtain bounds for the optical spectral weight for flat band superconductors, 
that lead to upper bounds for the superfluid stiffness and the 2D $T_c$.
We focus on on-site attraction \rev{$|U|$}
on the Lieb lattice with trivial flat bands and on the $\pi$-flux model with topological flat bands.
For trivial flat bands, the low-energy optical spectral weight 
\rev{
$\widetilde{D}_\text{low} \leq \widetilde{n} |U| \Omega/2$ with $\widetilde{n} = \min\left(n,2-n\right)$, where $n$ is the 
flat band density} 
and $\Omega$ the Marzari-Vanderbilt spread of the Wannier functions (WFs).
We also obtain a lower bound involving the quantum metric.
For topological flat bands, with an obstruction to localized WFs respecting all symmetries, we 
\rev{
again obtain an upper bound for $D_{\rm low}$ linear in $|U|$.} 
We discuss the insights obtained from our bounds by comparing them with mean-field and quantum Monte-Carlo results.
\end{abstract}

\maketitle

{\bf Introduction}: 
Understanding superconductivity (SC) in flat band systems is an important problem that has come to prominence with recent experiments on twisted bilayer~\cite{CaoNature2018a,CaoNature2018, BalentsNatPhys2020} and trilayer graphene \cite{ParkNature2021}.
The theoretical challenges are two-fold. First, the interactions are comparable to, or much larger than, the bandwidth, making
it a strong coupling problem.
Second, the topological nature of the bands
\cite{WatanabePNAS2015,PoNatComms2017,BradlynNature2017,PoPRX2018,PoPRB2019,SongPRL2019} acts as an obstruction for finding localized Wannier functions~\cite{MarzariPRB1997,MarzariRMP2012,BrouderPRL2007}.

Much of the recent progress, starting with the earliest proposals of flat band SC \rev{ \cite{KopninPRB2011}}
has been made using mean-field theory (MFT)~\rev{ \cite{TormaNatPhy2015,JulkuPRL2016,XiePRL2020,HuPRL2020,JulkuPRB2020,WangPRB2020}} 
and \rev{a variety of numerical methods for models with attractive~\cite{HofmannPRB2020,Aoki2020, PeriPRL2021} or repulsive interactions~\cite{KobayashiPRB2016, MondainiPRB2018}.}
The superfluid stiffness calculated within MFT is directly related to the 
the quantum geometry of the flat band \cite{TormaNatPhy2015}, however, the validity of the approximations used is not {\it apriori} clear in 
the strong coupling regime.
It is important to understand how interactions lead to transport and superconductivity
in a completely flat band, where the electrons have no kinetic energy.
Clearly, it would be very useful to have exact results which give insights and against which one can benchmark calculations
on strongly correlated problems involving topological bands.

In this paper, we derive exact bounds on the optical spectral weight 
and superfluid or phase stiffness $D_s$ of flat band SCs valid in any dimension. 
Our primary interest is in 2D, where these results yield upper bounds on the 
Berezinskii-Kosterlitz-Thouless $T_c$. We have earlier obtained very general results for multi-band Hamiltonians with
arbitrary interactions~\cite{HazraPRX2019}. Those bounds, though valid, are not very useful in the regime where 
only the flat bands are involved in the superconductivity, i.e.,
when the chemical potential lies in the flat bands and the interactions are smaller than the gap separating these bands from other states.
The problem is that multi-band spectral weight $\widetilde{D}$ of ref.~\cite{HazraPRX2019} includes {\it inter}-band transitions and thus
gives a gross overestimate of $D_s$ and $T_c$. Our goal is to compute low-energy  $\widetilde{D}_{\rm low}$ 
that captures only the {\it intra}-band spectral weight, and use it to obtain tight bounds on $D_s$ and $T_c$ for flat band SC.

Our main results are bounds on $\widetilde{D}_{\rm low}$ that scale linearly with the strength of on-site attraction $|U|$ 
\rev{
and depend on the flat band electron density $n$ via $\min\left(n,2-n\right)$}. 
They also depend crucially on the nature of the flat band Wannier functions (WFs). 
For topologically trivial flat bands, we find an upper bound on $\widetilde{D}_{\rm low}$ that involves the
Marzari-Vanderbilt (MV) spread $\Omega$~\cite{MarzariPRB1997} of the flat band WFs; see Eq.~\eqref{u-bound-1}.
We also obtain a lower bound on $\widetilde{D}_{\rm low}$ 
at $T\!=\!0$, provided the ground state exhibits off-diagonal long range order (ODLRO),
as has been established for some flat band Hamiltonians~\cite{Tasaki2020,ShenPRL1993}.
Our lower bound Eq.~\eqref{l-bound} is expressed in terms of the Fubini-Study quantum metric~\cite{MarzariPRB1997},
which shows that the quantum geometry of the flat band  
guarantees its non-zero interaction-induced optical spectral weight.

For topological flat bands, we consider a Kramers pair of spin $\uparrow$ and $\downarrow$ flat bands
with Chern numbers ${\cal C}_{\uparrow/\downarrow} = \pm 1$.
Naively one might think that $\Omega$ diverges, given the topological obstruction to obtaining 
exponentially localized WFs. However, following ref.~\cite{SoluyanovPRB2011,SoluyanovPRB2012},
one can construct localized WFs provided we accept that they are no longer Kramers pairs. Using these, we can find an upper bound that involves a ``generalized second moment'' of the WFs that is finite; see Eq.~\eqref{dt-low-pf}.

We illustrate the usefulness of our bounds for the attractive Hubbard models on two lattices:
(i) the Lieb lattice~\cite{Tasaki2020} which has topologically trivial flat bands, and 
(ii) the $\pi$-flux lattice~\cite{NeupertPRL2011},
whose dispersion can be tuned to harbor a Kramers pair of nearly flat topological bands.
We discuss the insights obtained from our exact results 
by comparing them with MFT on both models and recent 
quantum Monte Carlo data on the $\pi$-flux model~\cite{HofmannPRB2020}.
Our exact results validate the qualitative insight from MFT that flat band SC is closely tied to 
quantum geometry and nature of the WFs; in addition, they highlight the subtle but important differences between 
SC in trivial and topological flat bands.

{\bf Hamiltonian}: 
We start with a general Hamiltonians of the form $\mathcal{H} =  \mathcal{H}_K + \mathcal{H}_{\text{int}}$,
where the kinetic energy $\mathcal{H}_K =  \sum_{ {\bf i} {\bf j} \alpha\beta} t_{\alpha\beta}( {\bf i}- {\bf j} )
d^\dag_{ {\bf i} \alpha} d^{\phantom\dag}_{ {\bf j} \beta}$
describes hopping between orbitals $\alpha$ in a unit cell labelled by the Bravais lattice site ${\bf i}$. 
To begin with, we absorb spin into the orbital label $\alpha$.
The interaction $\mathcal{H}_{\text{int}} =
\sum_{ {\bf i}_j \alpha_j} V_{ 1234 }( {\bf i}_j, \alpha_j ) d^\dag_{ {\bf i}_1 \alpha_1} d^{\phantom\dag}_{ {\bf i}_2 \alpha_2} d^\dag_{ {\bf i}_3 \alpha_3} d^{\phantom\dag}_{ {\bf i}_4 \alpha_4}$. 

We analyze systems where the spectrum of $\mathcal{H}_K$ exhibits a very narrow band of bandwidth $w$, separated from
other dispersive bands by an energy gap $E_0 \gg w$ focusing primarily on the flat band limit $w=0$.  
(The transformation from orbital to band basis is described in appendix \ref{app-notation}.)  
The chemical potential lies within the flat band. 
The external vector potential  ${\bf A}$ couples to our system through a standard Peierls' substitution
in the kinetic energy $\mathcal{H}_K$, but does not couple to $\mathcal{H}_{\text{int}}$ in the the multi-band Hamiltonian. 
We discuss in detail below how ${\bf A}$ must couple to the interactions after we project down to flat bands.

We keep the discussion general to begin with, but
then specialize to two specific models, one with topologically trivial flat flat bands
and the other with nearly flat topological bands.
In both cases, we look at superconductivity arising from the simplest 
$V_{1234}$, an on-site attractive interaction of strength $|U| \gg w$
which induces singlet $s$-wave pairing.

{\bf Optical Spectral Weight}:
Our goal is to obtain exact results, without making mean field approximations or using numerical methods on finite latices,
for the integrated spectral weight $\widetilde{D} = ({\hbar^2}/{2 \pi e^2}) \int_0^\infty d\omega~{\rm Re}~\sigma(\omega)$,
an important observable in its own right. In a superconductor, $\widetilde{D}$ provides an upper bound~\cite{HazraPRX2019} on 
the superfluid stiffness $D_s$ and in 2D it serves to constrain the BKT $T_c$. 
Since $\sigma(\omega)$ is a tensor, we define $\widetilde{D}$ using the normalized trace 
$(1/2)\sum_\mu\sigma_{\mu\mu}(\omega)$ with $\mu = x,y$.
We focus on 2D, noting that the generalization to 3D is straightforward.
Coupling an external vector potential ${\bf A}$ to $\mathcal{H}$, we find that
$\widetilde{D} = (\hbar^2/8e^2 N_c) \langle \partial^2 \mathcal{H}/\partial A_\mu^2 \rangle$,
with an implicit sum on $\mu$. 

For $w \ll E_0 \lesssim |U|$ we need to use the full multi-band  $\mathcal{H}$, in which
${\bf A}$ couples only to $\mathcal{H}_K$ via the Peierls' phase factor, with the result~\cite{HazraPRX2019}
\begin{equation}
    \widetilde{D} = {1 \over{8 N_c } }\sum_{\bf k}
U _{m\alpha}({\bf k}){{\partial^2 t_{\alpha\beta}({\bf k})}\over{\partial k_\mu^2}}U^\dag_{\beta m^\prime}({\bf k})
\langle c^\dag_{{\bf k}m}c^{\phantom\dag}_{{\bf k}m'} \rangle.
\label{tildeD-full}
\end{equation}
\rev{Here, and below, the ${\bf k}$ sum is over the first Brillouin zone}, $N_c$ is the number of unit cells, and 
we implicitly sum over all repeated indices: bands  $m,m'$, orbitals $\alpha,\beta$, {\em and} space $\mu$.
The $U$-matrices diagonalize the orbital-space hopping to band energies:
\rev{
\begin{equation}
U_{m\alpha}({\bf k}) t_{\alpha\beta}({\bf k}) U^\dag_{\beta m^\prime}({\bf k}) = \epsilon_m({\bf k})\delta_{m,m'}.
\end{equation}}
Eq.~\eqref{tildeD-full} is a general multi-band result with $\langle c^\dag_m c^{\phantom\dag}_{m'} \rangle$  
defined by the fully interacting thermal expectation value.

Our primary interest is in the regime $w \ll |U| \ll E_0$, where Eq.~\eqref{tildeD-full} is not useful.
It includes {\it inter}-band transitions and gives an overestimate of the superfluid stiffness $D_s$ and thus $T_c$.
What we need is a low-energy $\widetilde{D}_{\rm low}$ that captures only the {\it intra}-band spectral weight,
and gives a tighter upper bound for $D_s$ and the 2D $T_c$.
\rev{Toward this end, we derive below an effective low-energy Hamiltonian $\mathcal{H}_{\rm low}$ that 
describes the system at energies below a cut-off $\Lambda$ to obtain
\begin{equation}
\widetilde{D}_{\rm low} = {{\hbar^2}\over{2 \pi e^2}} \int_0^\Lambda d\omega~{\rm Re}~\sigma(\omega) 
= {{\hbar^2}\over{8e^2 N_c}} \left\langle {{\partial^2 \mathcal{H}_{\rm low}}\over{\partial A_\mu^2}} \right\rangle.
\label{D-low}
\end{equation}
}
Following ref.~\cite{HazraPRX2019} we obtain bounds on the superfluid stiffness
$D_s \leq \widetilde{D}_{\rm low} < \widetilde{D}$, valid at each temperature and in all dimensions.
In 2D, using the BKT result $k_B T_c = \pi D_s(T_c^-)/2$, we further obtain $k_B T_c \leq \pi \widetilde{D}_{\rm low}/2$.
Thus we can use the $T$-independent upper bounds on $\widetilde{D}_{\rm low}$ that we derive below to bound the 2D $T_c$.

{\bf Low-energy Projected Hamiltonian}:
\rev{
We determine the low-energy Hamiltonian $\mathcal{H}_{\text{low}}$ in two steps. First, we
project to the subspace of (partially filled) flat bands labelled by $\ell$
focusing on the simplest case with a single pair of time-reversal partners in $\{ \ell \}$
that are separated from filled/empty bands by a gap $E_0$.
Later we will further project down to the low-energy many-body states within this subspace.
For the first step we use a standard technique, used for similar problems~\cite{HuberPRB2010,TovmasyanPRB2016}
and also for Chern insulators~\cite{Parameswaran2013}, which
essentially amounts to restricting the Hamiltonian to the subspace $\{ \ell \}$ of active bands when 
$E_0$ is the largest scale in the problem; see appendix \ref{app-projection}.
In the absence of kinetic energy ($w=0$), the low-energy effective Hamiltonian
consists of four fermion terms with coefficients 
$\widetilde{V}_{1234}( {\bf i}_j \ell_j ) = \sum_{ {\bf i}_j^\prime \alpha_j } V_{1234}( {\bf i}^\prime_j \alpha_j )
W_{\ell 1 \alpha_1}({\bf i}_1- {\bf i}_{ 1^\prime}) W^*_{\ell 2\alpha_2}({\bf i}_2- {\bf i}_{ 2^\prime}) W_{\ell_3\alpha_3}({\bf i}_3- {\bf i}_{ 3^\prime }) 
W^*_{\ell_4\alpha_4}({\bf i}_4- {\bf i}_{4^\prime})$.
Here $W_{\ell \alpha}({\bf r})$ are the Wannier functions related to $U_{\ell \alpha}({\bf k})$ via
\begin{equation}
W_{m\alpha}({\bf r}) =  \dfrac{1}{N_c} \sum\limits_{ {\bf k} } e^{ -i {\bf k} \cdot {\bf r} } U_{m\alpha}( {\bf k}).
\end{equation}
We must next reckon with the localization properties of the WFs, which depend crucially on 
whether the bands are trivial or topological~\cite{BrouderPRL2007,Monaco2018}.
}

{\bf Topologically Trivial Bands}:
We first analyze the simpler case of ``trivial'' bands, where we are guaranteed localized WFs~\cite{Monaco2018,BrouderPRL2007}
and thus we have a $\mathcal{H}_{\text{low}}$  with exponentially decaying interactions.
We shall see that the best upper bound for $\widetilde{D}_{\rm low}$ is obtained by choosing maximally
localized WFs of Marzari  and Vanderbilt (MV)~\cite{MarzariPRB1997,MarzariRMP2012}.

We use gauge invariance to determine how a uniform external vector potential ${\bf A}$ couples to $\mathcal{H}_{\text{low}}$ when we
have localized WFs; see appendix \ref{app-vecPotLatModel}. The interaction term  $\widetilde{V}_{ 1234 }( {\bf i}_j, \alpha_j)$ 
picks up a multiplicative phase factor of
 $\exp\left[ -i e({\bf r}_{ {\bf i}_4 \alpha_4} - {\bf r}_{ {\bf i}_3 \alpha_3} + {\bf r}_{ {\bf i}_2 \alpha_2} - {\bf r}_{ {\bf i}_1 \alpha_1})\! \cdot\!{\bf A} \right]$,
where ${\bf r}_{ {\bf i} \alpha}\!=\!{\bf i} + \tau_\alpha$ with $\tau_\alpha$ the location of orbital $\alpha$ in the unit cell.
We thus see that $\mathcal{H}_{\text{low}}$ couples to ${\bf A}$, even though the (e.g., density-density or spin-spin) interactions in the 
original multi-band $\mathcal{H}_{\text{int}}$ did not couple to ${\bf A}$. 

From this point onwards, we focus on the attractive Hubbard interaction in the multi-band Hamiltonian
$\mathcal{H}_\text{int} = - |U| \sum_{ {\bf i} \alpha } \hat{n}_{ {\bf i} \alpha \uparrow}  \hat{n}_{ {\bf i} \alpha \downarrow}$.
Here $\hat{n}_{ {\bf i} \alpha \sigma} = d^\dag_{ {\bf i} \alpha\sigma} d^{\phantom\dag}_{ {\bf i} \alpha\sigma}$ and
(from now on) the spin label $\sigma$ is explicit. \rev{
(For a generalization to orbital-dependent attractive interactions see appendix \ref{app-proj-D}).
}
We project this interaction down to the low-energy subspace spanned by
flat band eigenstates $\{ | \ell\!\uparrow\!{\bf k}\rangle, | \ell\!\downarrow\!{\bf k}\rangle\}$, and use
$W_{ \ell\uparrow, \alpha\uparrow }({\bf r}) = [W_{ \ell\downarrow, \alpha\downarrow }({\bf r})]^* \equiv W_{\ell\alpha}({\bf r})$
which follows from time-reversal. 

We generate a variety of terms in $\mathcal{H}_{\text{low}}$ all of which seem to be ${\cal O}(|U|)$, however,
the localization of the WFs allows us to organize these terms; 
\rev{
see appendix \ref{app-si-4}.}
The largest term is renormalized Hubbard attraction 
\rev{$|\widetilde{U}| = |U| \sum_{ {\bf i}^\prime \alpha} | W_{\ell\alpha}({\bf i}^\prime)|^4 $ }
that involves four WFs centered at the {\em same} site. 
\rev{
All other terms necessarily involve the overlap of WFs centered at {\em different} sites.
The characteristic length scale for WF decay decreases with increasing $E_0$, and we find that 
all other terms $\ll |\widetilde{U}|$ for large $E_0$, the regime where the projection is controlled.
}

The dominance of the on-site $|\widetilde{U}|$ over all other terms implies that we can further restrict the many-body Hilbert space
to states with only doubly occupied sites ($\uparrow\downarrow$) or empty sites,
\rev{
the attractive interaction analog of the ``lower Hubbard band''.
}
The low-energy degrees of freedom can be thought of as bosons, on-site pairs bound by the large attractive $|\widetilde{U}|$, with
a hard-core repulsion due to Pauli exclusion of constituent fermions. 
\rev{
Their kinetic energy in the low-energy subspace is governed by the
{\it pair hopping} amplitude $\mathcal{K}_{ij} = |U| \sum_{{\bf i}^\prime \alpha} | W_{\ell\alpha}({\bf i}-{\bf i}^\prime)|^2 | W_{\ell\alpha}({\bf j}-{\bf i}^\prime)|^2$.
The low-energy effective Hamiltonian is thus given by
\begin{equation}
    \mathcal{H}_{\text{low}}\!=\!-|\widetilde{U}| \sum\limits_{ {\bf i} } c^{\dag}_{ {\bf i} \ell \uparrow } c^{\phantom\dag}_{ {\bf i} \ell \uparrow } c^{\dag}_{ {\bf i} \ell \downarrow } c^{\phantom\dag}_{ {\bf i} \ell \downarrow }\! -\!  \sum\limits_{ \langle {\bf i},{\bf j}\rangle } \mathcal{K}_{ij} c^{\dag}_{ {\bf i} \ell \uparrow } c^{\dag}_{ {\bf i} \ell \downarrow } c^{\phantom\dag}_{ {\bf j} \ell \downarrow } c^{\phantom\dag}_{ {\bf j} \ell \uparrow} + \ldots.
\label{h-low}
\end{equation}
The ellipses denotes terms that do not couple to the vector potential
and thus do not contribute to ${{\partial^2 \mathcal{H}_{\rm low}}/{\partial A_\mu^2}}$.

Let us reiterate how $\mathcal{H}_{\text{low}}$ was derived. First, we projected to the flat-bands $\ell$ separated from
other bands by $E_0\!\gg\!|U|\!\gg\!w\!=\!0$. Second, we projected to the no single-occupancy subspace,
given that pair hopping $\mathcal{K}_{ij} \ll |\widetilde{U}|$ for localized WFs in the large $E_0$ regime.
Thus $\mathcal{H}_{\rm low}$ describes the physics within the ``lower Hubbard band'' with 
bandwidth set by the near-neighbor (NN) $\mathcal{K}_{ij}$, and in particular descibes its 
intraband optical spectral weight.
}

{\bf Spectral Weight Bound}:
\rev{
The vector potential ${\bf A}$ couples to $\mathcal{H}_{\text{low}}$ via 
$\mathcal{K}_{ij} \rightarrow \mathcal{K}_{ij} \exp \left[ i 2e( {\bf i} - {\bf j})\! \cdot\!{\bf A} \right]$ as explained above.
Using Eq.~\eqref{D-low} we obtain
}
\begin{equation}
\widetilde{D}_\text{low}  = \dfrac{|U|}{2} \sum\limits_{ {\bf i}, {\bf j} } \mathcal{D}_{ij}  P({\bf i}\!-\!{\bf j}),
\label{D-low_eq}.
\end{equation}
where $P({\bf i}\!-\!{\bf j})  =  \left\langle  c^{\dag}_{ {\bf i} \ell \uparrow } c^{\dag}_{ {\bf i} \ell \downarrow } c^{\phantom\dag}_{ {\bf j} \ell \downarrow } 
 c^{\phantom\dag}_{ {\bf j} \ell \uparrow } \right\rangle$ 
\rev{is the pairing correlation function} and 
$\mathcal{D}_{ij} = ({\bf i}\!-\!{\bf j})^2 \sum_{\alpha} | W_{\ell\alpha}({\bf i})|^2 | W_{\ell\alpha}({\bf j})|^2$.
\rev{
We show that $|P({\bf i}\!-\!{\bf j})| \leq \widetilde{n}/2$, where
$\widetilde{n} = \min\left(n,2-n\right)$, using the Cauchy-Schwarz 
inequality on a suitably defined inner product for operators; see appendix \ref{app-proj-D}.
We thus obtain the bound
 $\widetilde{D}_\text{low}  \leq (\widetilde{n} |U|/4) \sum_{ {\bf i}, {\bf j} } \mathcal{D}_{ij}$
 }
 
 We next relate this to the MV spread~\cite{MarzariPRB1997,MarzariRMP2012} for the
 flat band WFs $\Omega = \langle {\bf r}^2 \rangle_\ell - \langle {\bf r} \rangle_\ell^2$, where 
 we define $\langle f({\bf r}) \rangle_\ell \equiv \sum_{ {\bf r}, \alpha } f({\bf r})  |W_{\ell \alpha }( {\bf r} )|^2$. 
 We write
$\Omega = \left( \mathcal{D} + \mathcal{O}\right)/2$, with the orbital off-diagonal term
$\mathcal{O} = \sum_{ {\bf i},{\bf j}, \alpha \neq \beta} ({\bf i} - {\bf j})^2 |W_{ \ell \alpha }( {\bf i} )|^2 |W_{ \ell \beta }( {\bf j} )|^2$
and the orbital diagonal term can be expressed as $\mathcal{D} =  \sum_{ {\bf i},{\bf j}} \mathcal{D}_{ij}$.
Using $\mathcal{O} \geq 0$ we obtain
\begin{equation}
\widetilde{D}_\text{low} \leq \widetilde{n} |U| \Omega/2.
\label{u-bound-1}
\end{equation}
\rev{
The density dependence of bound, with $\widetilde{n} = \min\left(n,2-n\right)$, is sensible with the spectral weight
vanishing for both an empty and a filled flat band.
We note that $\widetilde{D}_\text{low}$ is in general a function of $T$ while our upper bound is $T$-independent
(within the low energy subspace). 
}

\begin{figure}
\includegraphics[scale=1]{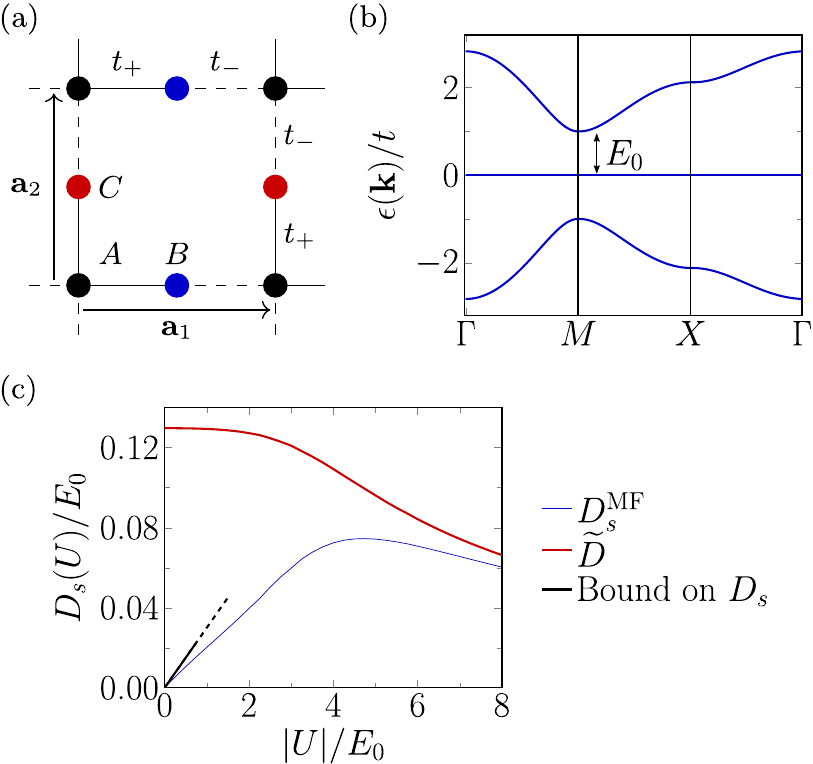}
\caption{
(a). Lieb lattice geometry: $A$ (black) orbitals lie on a square lattice, $B$ (blue) and $C$ (red) orbitals occupy bond centers. 
The staggered hopping with $t_\pm = t(1 \pm \delta)$, $0<\!\delta\!< 1$, isolates the flat band with a gap $E_0 = 2\sqrt{2}\delta t$.
(b). The band structure with $\Gamma = (0,0)$, $M = (\pi,\pi)$ and $X = (\pi,0)$. 
(c). The low energy spectral weight $\widetilde{D}_\text{low}$ bound on $D_s$ (black) derived by projecting to the flat band,
compared with the $T=0$ mean field superfluid stiffness $D_s^\text{MF}$ (blue) at half-filling, with $E_0 = t$.
The full three-band $\widetilde{D}$ (red) is calculated within MFT.
}
\label{fig-LiebLattice}
\end{figure}

{\bf Lieb lattice}:
We consider the attractive Hubbard model on the Lieb lattice
when the chemical potential $\mu$ is within the flat band. 
The kinetic energy
\begin{equation}
   \mathcal{H}_K = \sum\limits_{ {\bf k} \sigma} \Psi_{ {\bf k} }^\dag t({\bf k}) \Psi_{ {\bf k} }^{\phantom\dag}, \quad t({\bf k}) = \begin{pmatrix}
    0 & f^*(k_x) & f(k_y) \\
    f(k_x) & 0 & 0 \\
    f^*(k_y) & 0 & 0 
    \end{pmatrix} 
\label{lieb-blochH}
\end{equation}
describes nearest neighbor hopping on the Lieb lattice with three sites ($A,B,C$) per unit cell, with 
$\Psi_{ {\bf k} } = \left( d_{ {\bf k} A }, d_{ {\bf k} B }, d_{ {\bf k} C } \right)^T$
and $f(k) = t\left[ (1+\delta) + (1- \delta) e^{ i k } \right]$.
\rev{
The single-particle dispersion (see Fig.\ \ref{fig-LiebLattice})
has an exactly flat band at zero energy (${\rm det}[t({\bf k})] = 0$). 
This topologically trivial flat band is 
separated from two particle-hole symmetric bands by a 
gap $E_0 = 2\sqrt{2}\delta t$ controlled by $\delta$.
}

The flat band WFs have support only on the B and C sub-lattices, and mirror symmetry along 
the diagonal implies $\sum_{\bf i}|W_{\ell B}({\bf i})|^2 =  \sum_{\bf i}|W_{\ell C}({\bf i})|^2 = 1/2$. 
Using this we find $\Omega = \mathcal{D} + \varepsilon$ where 
$\varepsilon = \left[\sum_{\bf i}{\bf i}\, |W_{\ell B}({\bf i})|^2 -  \sum_{\bf i}{\bf i}\, |W_{\ell C}({\bf i})|^2\right]^2$;
see appendix \ref{app-Lieb}. Thus $\Omega \geq \mathcal{D}$, which allows us to improve Eq.~\eqref{u-bound-1} 
by a factor of two to obtain $\widetilde{D}_\text{low} \leq \widetilde{n} |U| \Omega/4$ for the Lieb lattice.

By an appropriate gauge choice in Eq.~\eqref{lieb-blochH} we can make $\varepsilon =0$ (see appendix \ref{app-Lieb} and \ref{app-gaugeBloch})
so that $\Omega =\mathcal{D}$.
This allows us to find an interesting lower bound on $\widetilde{D}_\text{low}$ at $T\!=\!0$.
Toward this end, note that Lieb's theorem~\cite{Tasaki2020,ShenPRL1993} has been used to establish 
off-diagonal long range order (ODLRO) at $T\!=\!0$
for the attractive Hubbard model on the Lieb latice when the flat-band is partially filled.
The \rev{pairing} correlation function (defined below Eq.~\eqref{D-low_eq}) $P({\bf r})$  
approaches $|\Phi|^2 \geq n(2-n)/36$ for $|{\bf r}|\to\infty$~\cite{Tasaki2020}.
We can view $|\Phi|^2 = n_{B0}$, the ``Bose condensate density''
for the projected $\mathcal{H}_{\text{low}}$ in Eq.~\eqref{h-low}, 
and we expect that $P({\bf r}) \geq n_{B0}$. Further using $\Omega = \mathcal{D}$,
Eq.~\eqref{D-low_eq} leads to
\begin{equation}
\widetilde{D}_{\text{low}}(T\!=\!0) \geq {n_{B0}\,|U|\over 2}\, \Omega > {n_{B0}\,|U| \over 2} {1\over N_c}\sum_{ {\bf k} }\text{Tr} \,g({\bf k}),
\label{l-bound}
\end{equation}
where $N_c$ is the number of unit cells in the system.

The second inequality in Eq.~\eqref{l-bound} emphasizes that  the 
spread $\Omega$ {\em cannot} be made arbitrarily small
by choice of WFs. The quantum metric $g({\bf k})$ (see appendix \ref{app-MVLocalization})
sets a lower bound~\cite{MarzariPRB1997} for the WF spread $\Omega$ and guarantees that the interaction-induced spectral weight is finite.
Thus our lower bound gives insight into the MFT result~\cite{JulkuPRL2016} showing 
the superfluid stiffness is directly related to the quantum geometry of the flat band WFs via $g({\bf k})$.
The inequality $N_c^{-1}\sum_{ {\bf k} }\text{Tr} \,g({\bf k}) \geq |{\cal C}|$ is not useful here since 
the Chern number ${\cal C}\!=\!0$ for a trivial band.

Next, we compare our exact results with $T\!=\!0$ MFT, focusing
on half-filling ($\mu = 0$). The calculation of the WFs and $\Omega$ is described in appendix \ref{app-Lieb}.
The MFT solution for the three-band
model is standard \cite{JulkuPRL2016,LiangPRB2017}.
This along with the calculation of three-band $\widetilde{D}$ and
stiffness $D_s$ within MFT is described in appendix \ref{app-mean-field-Ds}.

From Fig.~\ref{fig-LiebLattice}(c) we see that, in the regime $|U| \lesssim E_0 = 2\sqrt{2}\delta t$ where projection is justified,
our upper bound on $\widetilde{D}_{\text{low}} \geq D_s$ is linear in $|U|$, as is the
MFT result for $D_s$. (Note that MFT ignores quantum
fluctuations and will itself overestimate $D_s$.)
We can get further understanding of the slopes of these ``small $|U|$'' results as follows.
The current operator commutes with  $\mathcal{H}_{\text{low}}$,
the paramagnetic current susceptibility vanishes, and thus 
the diamagnetic term $\widetilde{D}_{\text{low}} = D_s$ at $T\!=\!0$. 
The WFs decay rapidly on the scale of the lattice spacing $a$ (see appendix \ref{app-Lieb})
and thus $P({\bf r})$ on for $r \approx a$ enters Eq.~\eqref{D-low_eq}.
The simple function 
 $P({\bf r}) \approx (\delta_{{\bf r},{\bf 0}} + 1)/4$
 interpolates on the scale of $a$ between $P({\bf 0}) = 1/2$ and 
$n_{B0} = 1/4$, the MFT ODLRO
at $T\!=\!0$ and half-filling ($n=1$).
This estimate gives insight into why our exact bound, which uses  $P({\bf 0})$, might 
overestimate the slope of $D_s$ by about a factor of two. 

Our lower bound \ref{l-bound} for $\widetilde{D}_{\text{low}}$
is also linear in $|U|$, but with a slope much smaller than MFT.
This arises because we
used the rigorous inequality~\cite{Tasaki2020} $n_{B0} \geq 1/36$ for ODLRO at half-filling,
while the MFT value is $1/4$. 

When $|U| \gtrsim E_0$, we cannot project down to the flat band, and
we must use the full three-band result Eq.~\eqref{tildeD-full}. One can use 
rigorous bounds~\cite{HazraPRX2019} for $\langle c^\dag_{{\bf k}m}c^{\phantom\dag}_{{\bf k}m'} \rangle$,
but in Fig.~\ref{fig-LiebLattice} we plot the MFT estimate of $\widetilde{D}$.
We also gain insight into the non-monotonic $|U|$-dependence of the superfluid stiffness $D_s$ from the perspective of the BCS-BEC crossover~\cite{RanderiaACMP2014}.
The ``small $|U|$'' regime is like a single-band BEC regime with a $D_s \sim |U|$
because the pair-hopping amplitude in Eq.~\eqref{h-low} scales like $|U|$.
This crosses over to a multi-band BEC at ``large $|U|$'' where $D_s \sim t^2/|U|$,
similar to the one-band attractive Hubbard model; see, e.g.,
Fig.~2 of ref.~~\cite{HazraPRX2019}.

\begin{figure}
\includegraphics[scale=1]{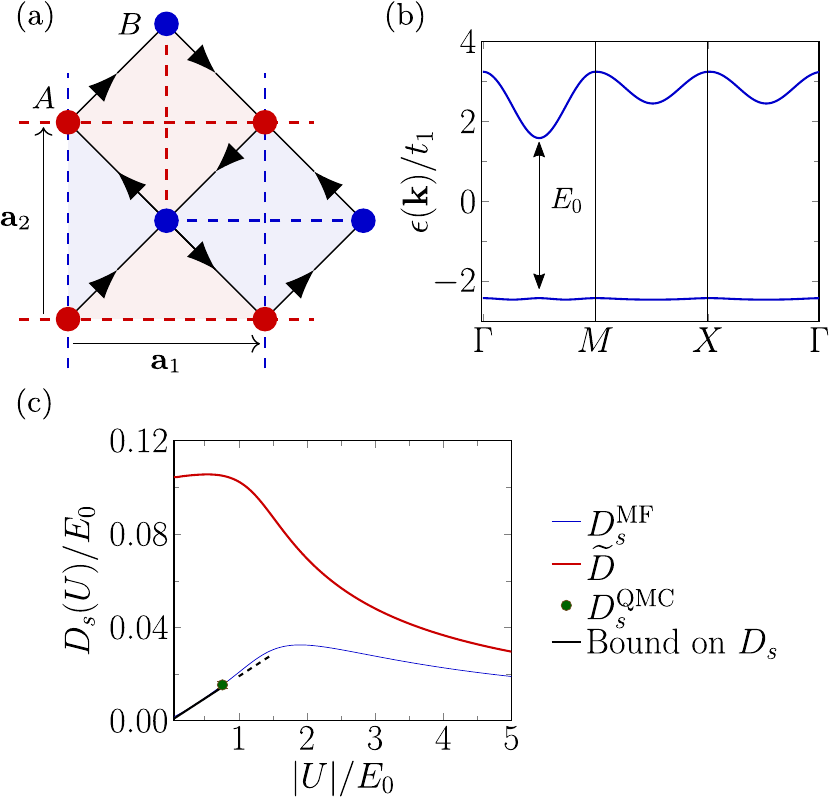}
\caption{
(a). $\pi$-flux model with $A$ (red) orbitals on a square lattice with $B$ (blue) orbitals at plaquette centers. 
Electrons hopping along the arrow pick up a phase $\sigma \pi/4$, where $\sigma = \pm$ labels the spin.
Dashed lines between NNN sites indicate hopping with opposite signs for blue and red labels. 
Fifth neighbor hopping (not shown) makes the lower band flat and isolates it with gap $E_0 = 3.94 t$.
(b). The band structure is shown along a high symmetry path.
(c). Comparison of upper bound (black line) from a projected calculation using spin-mixed localized Wannier functions,
with the $T=0$ mean-field $D_s^\text{MF}$ (blue) and two-band $\widetilde{D}$ (red) at half-filling of the nearly flat band.
The conventional (band curvature) contribution to the spectral weight of ${\cal O}(w)$ (estimated to be $0.008t$) is ignored; see text. 
We also show the QMC~\cite{HofmannPRB2020} result for $D_s$. 
}
\label{fig-PiFluxLattice}
\end{figure}

{\bf Topological Bands and $\pi$-Flux Model}: We finally turn to a model~
which harbors nearly-flat topological bands \cite{NeupertPRL2011} . It is a square-lattice analog of the Haldane model with two orbitals per unit cell
with near-neighbor (NN) hopping $t_1$ where spin $\sigma$ electrons hopping along the arrows in Fig.~\ref{fig-PiFluxLattice} 
pick up a phase $\pi\sigma/4$. The next-NN (NNN) hopping is chosen to be $t_2 = t_1/\sqrt{2}$ such that one obtains a particularly simple 
Bloch Hamiltonian ${\bf d}({\bf k})\!\cdot\!\boldsymbol\sigma$ (see appendix \ref{app-proj-weight-topo} for details). 
A  fifth-neighbor hopping adds a ${\bf d}_0({\bf k}) \mathbb{I}$ term and $t_5$ is tuned to flatten out the dispersion of the lower band 
(bandwidth $w \simeq 0.04 t_1 \ll E_0 \simeq 3.94 t_1$)
without impacting its non-trivial topology. 
The flat band consists of degenerate states $|\ell\!\uparrow{\bf k}  \rangle$ and  $| \ell\!\downarrow{\bf k} \rangle$ 
with Chern numbers ${\cal C}_{\uparrow/\downarrow} = \pm 1$.
Since $S_z$ is conserved, the  $\mathbb{Z}_2$ invariant for this subspace is
simply $\nu = ({\cal C}_{\uparrow} - {\cal C}_{\downarrow})/2$.

The Chern numbers are an obstruction for exponentially localized WFs which respect all the symmetries.
To circumvent this, we follow refs.~\cite{SoluyanovPRB2011,SoluyanovPRB2012}, and find a 
linear combination of $\uparrow$ and $\downarrow$ states, $|\ell 1 {\bf k} \rangle$ and  $| \ell 2 {\bf k} \rangle $, 
which are eigenstates of the Bloch Hamiltonian each with $\mathcal{C}=0$. 
The procedure for ``unwinding the topology'' is detailed in appendix \ref{app-proj-weight-topo}. Using these new states, we
obtain exponentially localized WFs $W_{\ell 1,\alpha\sigma}({\bf r})$ and $W_{\ell 2,\alpha\sigma}({\bf r})$;
the price we pay is that they do not form a Kramers pair. However, the WFs are sufficient for our purposes, especially since we do not make any approximations. 

We now find $\mathcal{H}_{\text{low}}$ in terms of the localized WFs,
which allows us to couple to a vector potential and obtain a finite $\widetilde{D}_{\rm low}$.
Unlike the Lieb lattice, we now have non-zero bandwidth $w$
of the nearly flat band which leads to a conventional 
\rev{contribution to} 
$\widetilde{D}_\text{low}$ of the form $\sum_k \partial^2\epsilon( {\bf k} ) n({\bf k})$. We focus on the regime
$w \ll |U| \ll E_0$, where this ${\cal O}(w)$ term is negligible.

In this regime $\mathcal{H}_{\text{low}}$ takes essentially the same form as Eq.~\eqref{h-low}
with the crucial difference that labels $\uparrow,\!\downarrow$ are replaced by the $1,2$ state.
The mathematical steps in obtaining and analyzing $\mathcal{H}_{\text{low}}$ are formally
similar to the trivial case since here too we now have exponentially localized WFs.
However, we should be careful since the $1,2$ states are not TR partners.
The pair hopping integral is no longer a density overlap (as it is in the trivial case; see above Eq.~\eqref{h-low}),
but has the form \rev{
$\mathcal{K}_{ij} =  |U|\sum_{{\bf i}^\prime \alpha\sigma} \big[ W_{\ell 1,\alpha\sigma}(\tilde{{\bf i}}) W_{\ell 2, \alpha\bar{\sigma}}(\tilde{{\bf i}}) 
W^*_{\ell 2, \alpha\bar{\sigma}}(\tilde{{\bf j}})  W^*_{\ell 1, \alpha\sigma}(\tilde{{\bf j}})  - W_{\ell 1,\alpha\sigma}(\tilde{{\bf i}}) W_{\ell 2, \alpha\bar{\sigma}}(\tilde{{\bf i}}) 
W^*_{\ell 1, \alpha\bar{\sigma}}(\tilde{{\bf j}})  W^*_{\ell 2, \alpha\sigma}(\tilde{{\bf j}}) \big] + ( \ell 1 \leftrightarrow \ell 2 )$ where $\tilde{{\bf i}} = {\bf i} + {\bf i}^\prime$, $\tilde{{\bf j}} = {\bf j} + {\bf i}^\prime$ and ${\bf i}^\prime$ is summed over.}
We see that $\mathcal{K}_{ij}$ is real and find that it is positive.

Following Eq.~\eqref{D-low_eq} and steps following it, we obtain  
\rev{
\begin{equation}
\widetilde{D}_\text{low} \leq \dfrac{\widetilde{n}}{4 N_c} \sum\limits_{ {\bf i}, {\bf j} } ( {\bf i} - {\bf j} )^2 |\mathcal{K}_{ij}|. 
\label{dt-low-pf}
\end{equation}
with $\widetilde{n} = \min\left(n,2-n\right)$.}
{\em If} $W_{\ell 1, \alpha\sigma}({\bf i}) = W^*_{\ell 2, \alpha\bar{\sigma}}({\bf i})$ {\em were true}, Eq.~\eqref{dt-low-pf} would 
reduce to the MV spread $\Omega$, however, this is {\em not true} for topological bands because of the 
topological obstruction. 

We compare in Fig.~\ref{fig-PiFluxLattice} our bounds with MFT (see appendix \ref{app-mean-field-Ds}) 
and QMC results~\cite{HofmannPRB2020} when the flat band is half-filled.
The $T\!=\!0$ MFT for the two-orbital attractive Hubbard model is similar to that for
the Lieb lattice, except that in addition to solving the two-gap equations, we need to 
self-consistently determine the chemical potential (since this model is not particle-hole symmetric). 

We see that the low-energy bound Eq.~\eqref{dt-low-pf}
captures the linear rise of $D_s$ in the small $|U|$ regime 
(with additional ${\cal O}(w)$ effects at very small $|U|$),
while the multi-band bound $\widetilde{D}$ goes to a constant at small $|U|$as it includes inter-band spectral weight.
The latter, obtained from the full two-band result Eq.~\eqref{tildeD-full} with 
$\langle c^\dag_m c^{\phantom\dag}_{m'} \rangle$ estimated using MFT, is 
of use in the $U\gtrsim E_0$ regime. The MFT result again shows the 
$t_1^2/|U|$ behavior of $\widetilde{D}$ and $D_s$ in the $U\gg E_0$ multiband-BEC regime.

It is an interesting open question to
understand why our bound Eq.~\eqref{dt-low-pf}, MFT and QMC
are so close to each other.
The agreement between the $T\!=\!0$ MFT superfluid stiffness $D_s$ and
the finite-$T$ QMC~\cite{HofmannPRB2020} could well be fortuitous.
Both methods likely overestimate $D_s$: MFT because of neglect of quantum fluctuations 
and QMC because of finite size effects.

{\bf Discussion}:
In conclusion,  we have presented exact bounds on the optical spectral spectral weight of flat band superconductors in
the strong coupling regime where the interaction is much larger than the bandwidth.
We obtained upper bounds on the superfluid stiffness and the 2D transition temperature that involve the geometry of the flat band Wannier functions.
\rev{
They scale linearly with the interaction $|U|$ and depend on the flat band electron density $n$ via $\min\left(n,2-n\right)$.}

For topologically trivial flat bands the upper bound is proportional to the MV spread of the WFs, while
for topological flat bands it is proportional to a generalized second moment of products of WFs.
The smaller the gap $E_0$ isolating the flat bands, the more the spatial spread of WFs, leading to  
a larger $D_s$ and $T_c$ for flat band SC.
We find, however, that if the gap $E_0$ becomes smaller than $|U|$,
multi-band effects begin to dominate~\cite{HazraPRX2019}
and ultimately lead to a drop in Ds and Tc.
Our main focus here has been flat band SC arising on-site attraction, but the methodology we have developed paves the way for analyzing strong correlations in flat band systems more generally,
including \rev{repulsive interactions and realistic models of twisted bilayer and trilayer graphene.}

{\bf Acknowledgements}:  This work was supported by NSF Materials Research
Science and Engineering Center (MRSEC) Grants
No. DMR-1420451 and No. DMR-2011876.

 
\bibliography{bounds-2}


\clearpage 

\onecolumngrid

\appendix

\begin{center}
{\bf APPENDIX}
\end{center}

\begin{enumerate}
\item[A.] Notation
\item[B.] Projection to low-energy subspace
\item[C.] Marzari-Vanderbilt Localization functional and Quantum metric
\item[D.] Low-energy Hamiltonian for Topologically Trivial Bands
\item[E.] Vector Potential in lattice models
\item[F.] Low-energy Spectral Weight
\item[G.] Lieb Lattice 
\item[H.] Projected Spectral Weight in Topological bands: $\pi$-flux model
\item[I.] Mean-field theory:
\begin{enumerate}
\item[1.] Gauge for the Multi-band Bloch Hamiltonian
\item[2.] Superfluid Stiffness
\item[3.] Hartree-Fock Corrections for Projected Hamiltonian
\end{enumerate}
\end{enumerate}

\renewcommand\thefigure{\thesection\arabic{figure}}

\section{Notation}\label{app-notation}

We start with a very general interacting, multi-orbital Hamiltonian of the form
\begin{eqnarray}
	\mathcal{H} =  \mathcal{H}_K + \mathcal{H}_{\text{int}} =  \sum\limits_{ {\bf i} {\bf j} \alpha\beta} t_{\alpha\beta}( {\bf i}- {\bf j} ) 
	d^\dag_{ {\bf i} \alpha} d^{\phantom\dag}_{ {\bf j} \beta} + \sum\limits_{ {\bf i}_j \alpha_j} V_{ 1234 }( {\bf i}_j, \alpha_j ) d^\dag_{ {\bf i}_1 \alpha_1} d^{\phantom\dag}_{ {\bf i}_2 \alpha_2} d^\dag_{ {\bf i}_3 \alpha_3} d^{\phantom\dag}_{ {\bf i}_4 \alpha_4} \label{Hfull-1-int}.
\end{eqnarray}
where ${\bf i}$ labels a unit cell, and $\alpha$ labels the orbitals within a unit cell. 
The locations of orbital $\alpha$ is ${\bf r}_{ {\bf i} \alpha} = {\bf i} + \tau_\alpha$, where $\{ \tau_\alpha \}$ is the basis set.
We find it convenient to absorb the spin degree of freedom into $\alpha$.
$\mathcal{H}_K$ defines a multi-orbital tight-binding model with hopping $t_{\alpha\beta}( {\bf i}- {\bf j} )$ 
from orbital $\beta$ in unit cell ${\bf j}$ to orbital $\alpha$ in unit cell ${\bf i}$.
$\mathcal{H}_{\text{int}}$ is a general four-fermion interaction (e.g., density-density or spin-spin), where we use the shorthand notation  
\begin{equation}
V_{ 1234 }( {\bf i}_j, \alpha_j ) \equiv V({\bf i}_1 \alpha_1, {\bf i}_2 \alpha_2, {\bf i}_3 \alpha_3, {\bf i}_4 \alpha_4).
\end{equation}

We transform from real-space ${\bf i}$ to reciprocal space ${\bf k} (\in $ first BZ) using
\begin{equation}
d^{\phantom\dag}_{ {\bf k} \alpha} = \dfrac{1}{\sqrt{N_c}} \sum\limits_{ {\bf i} } e^{ i {\bf k}\cdot {\bf i} } d^{\phantom\dag}_{ {\bf i} \alpha} \label{def-FT}
\end{equation}
where $N_c$ is the number of cells in the lattice. We thus obtain
\begin{eqnarray}
\mathcal{H}&=& \sum\limits_{ {\bf k} , \alpha\beta } t_{\alpha\beta}( {\bf k} ) d^{ \dag}_{ {\bf k} \alpha } d^{ \phantom\dag}_{ {\bf k} \beta} + \dfrac{1}{N_c} \sum\limits_{ {\bf k}_j \alpha_j } V_{ 1234 }( {\bf k}_j,\alpha_j) d^\dag_{ {\bf k}_1 \alpha_1} d^{\phantom\dag}_{ {\bf k}_2 \alpha_2} d^\dag_{ {\bf k}_3 \alpha_3} d^{\phantom\dag}_{ {\bf k}_4 \alpha_4}  \label{Hfull-2},
\end{eqnarray}
where
\begin{eqnarray}
t_{\alpha\beta}({\bf k}) &=& \sum\limits_{ {\bf r} } e^{ i {\bf k}\cdot {\bf r} } t_{ \alpha\beta}( {\bf r} ) \label{def-tab} \\
V_{ 1234 }( {\bf k}_j,\alpha_j)  &=& \dfrac{1}{N_c} \sum\limits_{ {\bf i}_j } e^{ i {\bf k}_1\cdot {\bf i}_1 - i {\bf k}_2\cdot {\bf i}_2 + 
i {\bf k}_3\cdot {\bf i}_3 - {\bf k}_4\cdot {\bf i}_4} \ V_{ 1234 }( {\bf i}_j, \alpha_j ) \label{def-vreal}.
\end{eqnarray}
The choice of Fourier transform in Eq.~\eqref{def-FT} results in a Bloch Hamiltonian that satisfies $t_{\alpha\beta}({\bf k}+{\bf G}) = t_{\alpha\beta}({\bf k})$ where $G$ is a reciprocal lattice vector. We find it convenient to work with this convention for finding Wannier functions.
Next, we use the unitary transformation $U_{m\alpha}({\bf k})$ from orbital basis $\alpha$ to band basis $m$
(including spin)
\begin{equation}
c_{ {\bf k} m} = \sum\limits_{\alpha} U_{m \alpha}({\bf k}) d_{ {\bf k} \alpha} \label{c-k-m-def}
\end{equation}
so that the kinetic energy is diagonalized
\begin{equation}
 \epsilon_m( {\bf k} ) = \sum\limits_{\alpha\beta} U^{\phantom\dag}_{m\alpha}( {\bf k}) t_{\alpha\beta}( {\bf k} ) U^\dag_{\beta m}( {\bf k}). \label{u-def}
 \end{equation}
 \rev{
 Note that at each ${\bf k}$ we are diagonalizing a $M \times M$ matrix where $M$ is the number of orbitals per unit cell, which also equals
 the number of bands.
 The matrix elements of the unitary transformation may be expressed as
 \begin{equation}
 U_{m\alpha}( {\bf k}) = \langle m {\bf k}\, \vert\, \alpha {\bf k} \rangle 
 \label{u-inner-product}
 \end{equation}
 a representation that will be useful later.
 } 
 
We can rewrite Eq.~\eqref{Hfull-2} in the band basis
\begin{equation}
\mathcal{H} =  \sum\limits_{ {\bf k}, m } \epsilon_m( {\bf k} ) c^{ \dag}_{ {\bf k} m } c^{ \phantom\dag}_{ {\bf k} m} + \dfrac{1}{N_c}\sum\limits_{ {\bf k}_j, m_j }  V_{ 1234 }( {\bf k}_j,m_j) c^{ \dag}_{ {\bf k}_1 m_1} c^{ \phantom\dag}_{ {\bf k}_2 m_2} c^{ \dag}_{ {\bf k}_3 m_3 } c^{ \phantom\dag}_{ {\bf k}_4 m_4} \label{Hfull-3}
\end{equation}
where
\begin{equation}
V_{ 1234 }( {\bf k}_j,m_j) = \sum\limits_{\{ \alpha \} } V_{ 1234 }( {\bf k}_j,\alpha_j) U^{\phantom\dag}_{m_1 \alpha_1 }( {\bf k}_1) U^\dag_{\alpha_2 m_2 }( {\bf k}_2 )  U^{\phantom\dag}_{m_3 \alpha_3 }( {\bf k}_3 ) U^\dag_{ \alpha_4 m_4}( {\bf k}_4) \label{def-Vell}.
\end{equation}

\section{Projection to the low-energy subspace}\label{app-projection}

We want to focus on the ``active bands'' $\ell \in \mathcal{L}$ the low-energy subspace. We assume that the chemical potential lies in the active bands, which are separated from all other -- either completely filled or completely empty -- bands by an 
energy scale $E_0$. Let $w$ be the effective hopping scale (or bandwidth) scale of the active bands, and $V$ be the scale of the interaction
terms. We next find the effective low-energy Hamiltonian projected to $\mathcal{L}$ 
when $w \ll E_0$ and $V \ll E_0$, and arbitrary $V/w$. Under these conditions, projection 
\begin{equation}
\mathcal{P} c_{ {\bf k} m } \mathcal{P} =  c_{ {\bf k} \ell } \delta_{m,\ell}
\label{c-proj}
\end{equation}
simply amounts to restricting the Hamiltonian to bands within $\mathcal{L}$.
\rev{
Such an approach is widely used in quantum Hall problems for projecting interactions to the lowest Landau level, and
has also been used for lattice problems of Chern insulators~\cite{Parameswaran2013}.
}
We thus find
\begin{equation}
\mathcal{H}_{\rm low} =  \sum\limits_{ {\bf k}, \ell \in \mathcal{L} } \epsilon_\ell( {\bf k} ) c^{ \dag}_{ {\bf k} \ell } 
c^{ \phantom\dag}_{ {\bf k} \ell} + \dfrac{1}{N_c}\sum\limits_{ {\bf k}_j, \ell_j \in \mathcal{L} }  V_{ 1234 }( {\bf k}_j,\ell_j)  
c^{ \dag}_{ {\bf k}_1 \ell_1} c^{ \phantom\dag}_{ {\bf k}_2 \ell_2} c^{ \dag}_{ {\bf k}_3 \ell_3 } c^{ \phantom\dag}_{ {\bf k}_4 \ell_4}
\end{equation}

\begin{figure}
\centering
\includegraphics[scale=1]{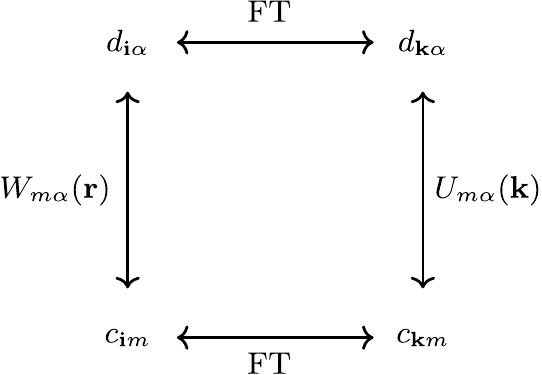}
\caption{Illustrating the relationship between different representations: Here ${\bf i}$ labels a unit cell in real space, ${\bf k}\in $ BZ,
$\alpha$ is an orbital label and $m$ is a band label. To begin with the spin label is absorbed into the $\alpha$ and $m$ labels;
later on we make it explicit.}
\label{fig1-notation}
\end{figure}

In order to understand how external EM fields couple to this Hamiltonian, we need a real space representation of $\mathcal{H}_{\rm low}$.
To achieve that, we use the Fourier transform
\begin{equation}
c^{\phantom\dag}_{ {\bf i} m} = \dfrac{1}{\sqrt{N_c}}\sum\limits_{ {\bf k} } e^{ -i {\bf k}\cdot {\bf i} } c^{\phantom\dag}_{ {\bf k} m} \label{c-i-m-def}.
\end{equation}
to introduce the operator $c^\dag_{ {\bf i} m}$. It creates an electron in the Wannier orbital of band $m$ in unit cell ${\bf i}$ \cite{MarzariRMP2012}.
We can relate the Wannier orbital to the original tight-binding orbitals
\begin{eqnarray}
c^{\phantom\dag}_{ {\bf i} m} =  \dfrac{1}{\sqrt{N_c}} \sum\limits_{ {\bf k} } e^{ -i {\bf k}\cdot {\bf i} } c^{\phantom\dag}_{ {\bf k} m} &=&  
\sum\limits_{ {\bf j}, \alpha } d^{\phantom\dag}_{ {\bf j} \alpha}  W_{m\alpha}( {\bf i}-{\bf j} ) \label{def-WO}
\end{eqnarray}
by defining the Wannier function
\begin{equation}
W_{m\alpha}({\bf r}) =  \dfrac{1}{N_c} \sum\limits_{ {\bf k} } e^{ -i {\bf k} \cdot {\bf r} } U_{m\alpha}( {\bf k}) \label{def-WF}.
\end{equation}
The relation between different bases is summarized in Fig.~\ref{fig1-notation}.

We are now ready to write the low-energy Hamiltonian in terms of the Wannier orbitals
\begin{equation}
\mathcal{H}_{\text{low}} = \sum\limits_{ {\bf i}_j,  \{ \ell_i \} \in \mathcal{L} } \tilde{t}_{ \ell_1 \ell_2 }({\bf i}_1-{\bf i}_2) c^{\dag}_{ {\bf i}_1 \ell_1} c^{\phantom\dag}_{ {\bf i}_2 \ell_2 } +  \sum\limits_{ {\bf i}_j,  \ell_j \in \mathcal{L} } \widetilde{V}_{1234}( {\bf i}_j, \ell_j ) c^\dag_{ {\bf i}_1  \ell_1} c^{\phantom\dag}_{ {\bf i}_2 \ell_2 } c^\dag_{ {\bf i}_3 \ell_3 } c^{\phantom\dag}_{ {\bf i}_4 \ell_4 }.
\label{H-low}
\end{equation}
where the real-space hopping and interaction matrix elements (denoted by `tilde') are given by the appropriate overlaps of Wannier functions
\begin{eqnarray}
\tilde{t}_{ \ell_1 \ell_2 }({\bf i}_1-{\bf i}_2) &=& \sum\limits_{ {\bf i}_j^\prime, \alpha\beta }  t_{\alpha \beta}( {\bf i}_{1^\prime} - {\bf i}_{2^\prime} ) W_{\ell_1 \alpha}({\bf i}_1- {\bf i}_{ 1^\prime}) W^*_{\ell_2 \beta}({\bf i}_2- {\bf i}_{ 2^\prime}) \label{def-tildet} \\
\widetilde{V}_{1234}( {\bf i}_j, \ell_j )&=& \sum\limits_{ {\bf i}_j^\prime , \alpha_j } V_{1234}( {\bf i}^\prime_j, \alpha_j )W_{\ell_1 \alpha_1}({\bf i}_1- {\bf i}_{ 1^\prime}) W^*_{\ell_2\alpha_2}({\bf i}_2- {\bf i}_{ 2^\prime}) W_{\ell_3\alpha_3}({\bf i}_3- {\bf i}_{ 3^\prime }) W^*_{\ell_4\alpha_4}({\bf i}_4- {\bf i}_{4^\prime}). \label{def-Vtilde}
\end{eqnarray}
Notice that the hopping matrix elements is in general off-diagonal in the band label.
When the low-energy subspace has band degeneracies or some topological character, one has to work with generalized Wannier orbitals.
They span the same Hilbert space $\mathcal{L}$ but do not have one-to-one correspondence with the original bands $\{ \ell_j \}$.

We emphasize that the operators $c^{\dag}_{ {\bf i} \ell}$ and $c^{\phantom\dag}_{ {\bf i} \ell}$ that enter $\mathcal{H}_{\text{low}}$
of Eq.~\eqref{H-low} obey standard fermion anticommutation relations as they are just the Fourier transforms of the
original fermion operators (see Eq.~\eqref{c-i-m-def}). 
We contrast this with other bases in which projected operators may obey non-trivial commutation relations. 
Let us, for instance, write the low-energy effective Hamiltonian in the orbital basis (rather the band basis chosen above).
We then invert Eq.~\eqref{def-WO}, using the orthonormality of Wannier functions, to obtain
\begin{equation}
d^{\phantom\dag}_{ {\bf i} \alpha} =  \sum\limits_{ {\bf j}, m } c^{\phantom\dag}_{ {\bf j} m}  W^*_{\alpha m}( {\bf i}-{\bf j} ).\label{d-i-a-def}
\end{equation}
This in turn can be projected to the low-energy subspace via
$\mathcal{P} d_{ {\bf i} \alpha} \mathcal{P} =  \tilde{d}^{\phantom\dag}_{ {\bf i} \alpha}$.
Using Eq.~\eqref{c-proj} we find
\begin{equation}
\tilde{d}^{\phantom\dag}_{ {\bf i} \alpha} = 
 \sum\limits_{ {\bf j} } c^{\phantom\dag}_{ {\bf j} \ell}  W^*_{\alpha \ell}( {\bf i}-{\bf j} ).
\label{tilde-def}
\end{equation}
These operators now obey non-trivial commutation relations 
\begin{equation}
\left\{  \tilde{d}^{\dag}_{ {\bf i} \alpha},  \tilde{d}^{\phantom\dag}_{ {\bf j} \beta} \right\} = \sum\limits_{ {\bf i}^\prime }  W_{\ell \alpha}( {\bf i}- {\bf i}^\prime )W^*_{\beta \ell}( {\bf j}- {\bf i}^\prime ).
\end{equation}

\section{Marzari-Vanderbilt Localization functional and Quantum Metric}\label{app-MVLocalization}

\rev{
The localization properties of Wannier functions play a crucial role in our analysis in various ways, including the derivation of the effective low-energy Hamiltonian, 
coupling an external vector potential to the system, and in the ultimate form of the bound that we derive. Here we briefly summarize those aspects of this 
topic that are directly relevant for our work, particularly the Marzari-Vanderbilt spread for WFs~\cite{MarzariPRB1997} and the quantum metric. For a full discussion, 
we refer the reader to the original litertature~\cite{MarzariPRB1997,MarzariRMP2012}.

The Bloch states have a gauge ambiguity $| \ell {\bf k} \rangle \longrightarrow \exp{ (i \phi({\bf k}) ) } | \ell {\bf k} \rangle$, and depending on the choice of these 
phases one can  greatly impact the localization properties of the  WFs which are the Fourier transforms of the Bloch wave functions.
Marzari and Vanderbilt~\cite{MarzariPRB1997} argued that the second moment of the Wannier function 
\begin{eqnarray}
\Omega =\langle {\bf r}^2 \rangle_\ell - |\langle {\bf r} \rangle_\ell|^2 &=& \sum\limits_{ {\bf i}, \alpha } {\bf i}^2  |W_{ \ell \alpha }( {\bf i} )|^2 - \sum\limits_{ {\bf i}, \alpha } \sum\limits_{ {\bf j}, \beta } ({\bf i} \cdot {\bf j})  |W_{ \ell \alpha }( {\bf i} )|^2 |W_{ \ell \beta }( {\bf j} )|^2 \label{def-Omega} .
\end{eqnarray}
should be minimized for the optimal choice of gauge.}
The Wannier functions thus obtained are called maximally localized Wannier functions (MLWF).
The spread functional is decomposed into two terms $\Omega = \Omega_I + \widetilde{\Omega}$ where
\begin{eqnarray}
\Omega_I &=&\sum\limits_{ {\bf i}, \alpha } {\bf i}^2  |W_{ \ell \alpha }( {\bf i} )|^2 - \sum\limits_{ {\bf i}, \alpha } \sum\limits_{ {\bf j}, \beta } ({\bf i} \cdot {\bf j}) \sum\limits_{ {\bf i}^\prime }  W_{  \alpha \ell }^\dag( {\bf i} ) W^{\phantom\dag}_{\ell\alpha}( {\bf i} + {\bf i}^\prime) W_{ \beta\ell}^\dag( {\bf j}- {\bf i}^\prime ) W_{\ell\beta}^{\phantom\dag}( {\bf j}) \\
\widetilde{\Omega} &=& \sum\limits_{ {\bf i}, \alpha } \sum\limits_{ {\bf j}, \beta }({\bf i} \cdot {\bf j}) \sum\limits_{ {\bf i}^\prime \neq {\bf 0} }   W_{  \alpha \ell }^\dag( {\bf i} ) W^{\phantom\dag}_{\ell\alpha}( {\bf i} + {\bf i}^\prime) W_{ \beta\ell}^\dag( {\bf j}- {\bf i}^\prime ) W_{\ell\beta}^{\phantom\dag}( {\bf j}) 
\end{eqnarray}
such that $\Omega_I$ is gauge-invariant and $\widetilde{\Omega}$ is a positive quantity that needs to be minimized.

We next make connections with the quantum metric $g_\ell({\bf k})$, the Berry connection $\mathcal{A}_\ell({\bf k})$ and the Berry curvature $f_\ell({\bf k})$ of band $\ell$.
We use the identity
\begin{equation}
\boldsymbol\nabla_{\bf k} U_{ \ell \alpha }({\bf k}) = i \sum\limits_{ {\bf i} } {\bf i} \; e^{ i {\bf k}\cdot {\bf i} } W_{\ell\alpha}({\bf i})
\end{equation}
to re-write
\begin{eqnarray}
\Omega_I &=& \dfrac{1}{N_c}\left( \sum\limits_{ {\bf k}\alpha } \boldsymbol\nabla_{ {\bf k} } U^\dag_{ \alpha \ell}({\bf k}) \cdot \boldsymbol\nabla_{ {\bf k} } U^{\phantom\dag}_{ \ell \alpha }({\bf k}) - \sum\limits_{ {\bf k}, \alpha\beta } U^\dag_{ \alpha \ell}({\bf k}) \left( \boldsymbol\nabla_{ {\bf k} } U^{\phantom\dag}_{ \ell \alpha }({\bf k})\right) \cdot \left(\boldsymbol\nabla_{ {\bf k} } U^{\dag}_{ \beta \ell }({\bf k}) \right) U^{\phantom\dag}_{ \ell \beta }({\bf k}) \right) = \dfrac{1}{N_c} \sum\limits_{{\bf k}} \text{Tr}[ g_\ell({\bf k}) ] \label{def-Omega-I}\\
\widetilde{\Omega} &=& \dfrac{1}{N_c} \sum\limits_{ {\bf k}, \alpha\beta } U^\dag_{ \alpha \ell}({\bf k}) \left( \boldsymbol\nabla_{ {\bf k} } U^{\phantom\dag}_{ \ell \alpha }({\bf k})\right) \cdot \left(\boldsymbol\nabla_{ {\bf k} } U^{\dag}_{ \beta \ell }({\bf k}) \right) U^{\phantom\dag}_{ \ell \beta }({\bf k})   - \left( \dfrac{1}{N_c} \sum\limits_{ {\bf k}, \alpha } U^\dag_{ \alpha \ell}({\bf k}) \boldsymbol\nabla_{ {\bf k} } U^{\phantom\dag}_{ \ell \alpha }({\bf k}) \right)\cdot \left( \dfrac{1}{N_c} \sum\limits_{ {\bf k}^\prime, \beta } \boldsymbol\nabla_{ {\bf k} }  U^\dag_{ \beta \ell}({\bf k}^\prime) U^{\phantom\dag}_{ \ell \beta }({\bf k}^\prime) \right)  \nonumber  \\
&=& \dfrac{1}{N_c} \sum\limits_{ {\bf k} } | \mathcal{A}_\ell({\bf k}) |^2 - \left| \dfrac{1}{N_c} \sum\limits_{ {\bf k} } \mathcal{A}_\ell \right|^2 = \dfrac{1}{N_c} \sum\limits_{ {\bf k} } | \mathcal{A}_\ell({\bf k}) - \bar{\mathcal{A}_\ell} |^2 \label{def-Omega-tilde}
\end{eqnarray}
Since $\widetilde{\Omega}$ is a sum of positive number, $\widetilde{\Omega}\geq 0$ and the trace of quantum metric sets a lower bound on the spread
\begin{equation}
\Omega \geq \Omega_I = \dfrac{1}{N_c} \sum\limits_{{\bf k}} \text{Tr}[ g_\ell({\bf k}) ]. \label{omega-quantum}
\end{equation}

The quantum metric and Berry curvature are the real and imaginary parts of the quantum geometric tensor $\mathcal{B}_\ell({\bf k})$
\begin{equation}
[\mathcal{B}_\ell({\bf k}) ]_{\mu\nu} = \sum\limits_{\alpha\beta} \partial_\mu U^\dag_{ \alpha \ell}({\bf k}) \left[ 1 - U^{\phantom\dag}_{ \ell \alpha } U^{\dag}_{ \beta \ell } \right] \partial_\nu U^{\phantom\dag}_{ \ell \beta }({\bf k}) = [g_\ell({\bf k})]_{\mu\nu} - \dfrac{i}{2} [f_\ell({\bf k})]_{\mu\nu}.
\end{equation}
\rev{
It is useful to write this in the bra-ket notation. Bloch's theorem states that the eigenstates are of the form
$|\ell {\bf k} \rangle = e^{ i {\bf k} \cdot \hat{{\bf r}} }\, |u_{\ell {\bf k}} \rangle$, where 
 $|u_{\ell {\bf k}} \rangle$ are the the cell-periodic Bloch states; note the position operator $\hat{{\bf r}}$ here.
The orbital states are given by
\begin{equation}
|\alpha {\bf k} \rangle = \dfrac{1}{\sqrt{N_c}} \sum\limits_{ {\bf i} } e^{ i {\bf k} \cdot {\bf i}} | \alpha {\bf i} \rangle
\end{equation}
where the function $\langle {\bf r} | \alpha {\bf i} \rangle$ is strongly localized in the unit cell ${\bf i}$, given the multi-band tight-binding Hamiltonian.
Thus $\hat{{\bf r}} | \alpha {\bf i} \rangle = {\bf i} | \alpha {\bf i} \rangle$ and we can rewrite
\begin{equation}
|\alpha {\bf k} \rangle = e^{ i {\bf k} \cdot \hat{{\bf r}}} \left(\dfrac{1}{\sqrt{N_c}} \sum\limits_{ {\bf i} } | \alpha {\bf i} \rangle \right) \equiv e^{ i {\bf k} \cdot \hat{{\bf r}}} | \alpha \rangle.
\end{equation}
We can then rewrite
\begin{equation}
U_{\ell \alpha}( {\bf k}) = \langle \ell {\bf k} | \alpha {\bf k} \rangle = \left( \langle u_{\ell {\bf k}} | e^{ -i {\bf k} \cdot \hat{{\bf r}}} \right) \left( e^{ i{\bf k} \cdot \hat{{\bf r}}} | \alpha \rangle \right) =  \langle u_{\ell {\bf k}} | \alpha \rangle
\end{equation}
from which we obtain
 \begin{equation}
 \partial_\mu U_{\ell \alpha}( {\bf k}) = \langle \partial_\mu u_{\ell {\bf k}} \vert\, \alpha \rangle 
 \end{equation}
where $\partial_\mu = \partial/\partial k_\mu$. From unitarity of the $U_{\ell \alpha}( {\bf k})$ matrix at each ${\bf k}$, we see that 
\begin{equation}
\sum\limits_\alpha U_{m \alpha}({\bf k}) U^\dag_{ \alpha n}({\bf k}) = \delta_{mn} \Rightarrow \sum\limits_\alpha \langle u_{ m {\bf k} } | \alpha \rangle \langle \alpha | u_{ n {\bf k}}  \rangle = \delta_{mn}
\end{equation}
hence, the states $\alpha$ form a complete set within the space spanned by $| u_{ \ell {\bf k}} \rangle$. Using this, we find
\begin{equation}
\sum\limits_{\alpha} \partial_\nu U^{\phantom\dag}_{ \ell \alpha }({\bf k}) \partial_\mu U^\dag_{ \alpha \ell}({\bf k}) = \langle \partial_\nu u_{\ell {\bf k}}  | \partial_\mu u_{\ell {\bf k}} \rangle, \quad \sum\limits_{\alpha} U^{\phantom\dag}_{ \ell \alpha }({\bf k}) \partial_\mu U^\dag_{ \alpha \ell}({\bf k})  = \langle u_{\ell {\bf k}}  | \partial_\mu u_{\ell {\bf k}} \rangle.
\end{equation}
Finally, we can write the quantum geometric tensor in a more illuminating form
\begin{equation}
[\mathcal{B}_\ell({\bf k}) ]_{\mu\nu} = \left\langle \partial_\nu u_{\ell {\bf k}} \left| \left(1 - |u_{\ell {\bf k}} \rangle\langle u_{\ell {\bf k}} | \right) \right| \partial_\mu u_{\ell {\bf k} } \right\rangle.
\end{equation}

Since $\mathcal{B}_\ell({\bf k})$ involves the complementary projector (in parenthesis), one of its eigenvalues is zero and its determinant vanishes.}
In 2D, that leads to the relation $\text{det}[ g_\ell({\bf k}) ] = |f_{\ell}({\bf k})|^2/4$. We can further invoke $2\sqrt{\text{det}[\cdot]} \leq \text{Tr}[\cdot]$ to arrive at $\text{Tr}[ g_\ell ({\bf k} )] \geq | f_{\ell}({\bf k}) |$ \rev{ and 
\begin{eqnarray}
\Omega_I = \dfrac{1}{N_c} \sum\limits_{{\bf k}} \text{Tr}[ g_\ell({\bf k}) ] \geq \dfrac{1}{N_c} \sum\limits_{{\bf k}} | f_{\ell}({\bf k}) | \geq \dfrac{1}{N_c} \left| \sum\limits_{{\bf k}} f_{\ell}({\bf k}) \right| = |\mathcal{C}_\ell|
\end{eqnarray}
where $\mathcal{C}_\ell$ is the Chern number. In that sense, Chern number sets a lower bound for the spread.
However, that is misleading since the other piece ($\widetilde{\Omega})$ diverges when $\mathcal{C}_\ell \neq 0$ and forces $\Omega \rightarrow \infty$ \cite{Monaco2018}.}

The origin of this divergence can be traced back to the smoothness of phase.
Chern numbers are an obstruction to a smooth gauge, and that obstruction manifests in $\widetilde{\Omega}$. We define a \emph{smoothness} function
\begin{equation}
\mathcal{S}_\ell({\bf k}) = \left| {\bf \mathcal{A}}_\ell({\bf k}) \right|^2 \label{def-smoothness}
\end{equation}
as a diagnostic tool to track smoothness of the phase.
There is a direct relation between $\mathcal{S}_\ell({\bf k})$ and $\widetilde{\Omega}$ as seen in Eq.~\eqref{def-Omega-tilde}.
$\mathcal{S}_\ell({\bf k})$ has singularities in the BZ (see Fig.~\ref{fig-pf-updown}) that make $\widetilde{\Omega}$ diverge.
Numerically, we track the divergence by computing $\Omega(R_c)$ where $R_c$ is a cutoff scale
\begin{equation}
\Omega(R_c) = \sum\limits_{  \alpha, |{\bf i}|\leq R_c } {\bf i}^2  |W_{ \ell \alpha }( {\bf i} )|^2 - \sum\limits_{\alpha, |{\bf i}| \leq R_c } \sum\limits_{ \beta, |{\bf j}| \leq R_c } ({\bf i} \cdot {\bf j})  |W_{ \ell \alpha }( {\bf i} )|^2 |W_{ \ell \beta }( {\bf j} )|^2. \label{omega-Rc}
\end{equation}
We examine $\Omega(R_c)$ as a function of $R_c$ to determine convergence.

\section{Low-energy Hamiltonian for Topologically Trivial Bands}\label{app-si-4}

We will focus on a class of systems that have one isolated band (per spin), say $\ell$, with no topological obstruction to exponential localization,
and derive the low-energy Hamiltonian.

Let us make the spin-label explicit.
The low energy Hilbert space is spanned by bands $\{ \ell \uparrow, \ell \downarrow \}$ where the spin sectors are independent and related by time reversal. In other words,
\begin{equation}
W_{ \ell \uparrow, \alpha \uparrow }({\bf r}) = [W_{ \ell \downarrow, \alpha \downarrow }({\bf r})]^* \equiv W_{\ell\alpha}({\bf r}), \quad \text{and} \quad W_{ \ell \uparrow, \alpha \downarrow }({\bf r}) = W_{ \ell \downarrow, \alpha \uparrow }({\bf r}) = 0
\end{equation}
These relations allow us to drop the spin label and focus on $W_{\ell\alpha}({\bf r})$.

\rev{
We focus here on the {\it attractive} Hubbard model with 
$\mathcal{H}_\text{int} = - |U| \sum_{ {\bf i} \alpha } \hat{n}_{ {\bf i} \alpha \uparrow}  \hat{n}_{ {\bf i} \alpha \downarrow}$.
The generalization to orbital-dependent attractive interactions is given below; see Eq.~\eqref{H-general}.}
We use Eq.~\eqref{def-Vtilde} to find the projected interactions
\begin{equation}
\widetilde{V}_{1234}( {\bf i}_j ) = - |U| \sum\limits_{ {\bf i}^\prime, \alpha } W_{\ell \alpha}({\bf i}_1- {\bf i}^\prime) W^*_{\ell \alpha}({\bf i}_2- {\bf i}^\prime) W^*_{\ell \alpha}({\bf i}_3- {\bf i}^\prime)  W_{\ell \alpha}({\bf i}_4- {\bf i}^\prime) \label{v1234-wf}
\end{equation}
and the associated projected Hamiltonian $\mathcal{H}_\text{proj} = \sum\limits_{ {\bf i}_j } \widetilde{V}_{1234}( {\bf i}_j ) c^\dag_{ {\bf i}_1  \ell\uparrow} c^{\phantom\dag}_{ {\bf i}_2 \ell \uparrow } c^\dag_{ {\bf i}_3 \ell\downarrow } c^{\phantom\dag}_{ {\bf i}_4 \ell\downarrow }$.

%
\rev{
There are many different choices of $\{ {\bf i}_i\}$, each of which leads to an interaction term of which seems to be of ${\cal O}(|U|)$.
In particular, there is: Renormalized Hubbard ${\bf i}_1 = {\bf i}_2 = {\bf i}_3 = {\bf i}_4$, Density-dependent hopping ${\bf i}_1 = {\bf i}_2 = {\bf i}_3 \neq {\bf i}_4$, Density-Density interaction ${\bf i}_1 = {\bf i}_2 \neq {\bf i}_3 = {\bf i}_4$, Pair hopping ${\bf i}_1 = {\bf i}_3 \neq {\bf i}_2 = {\bf i}_4$ and Spin-flip ${\bf i}_1 = {\bf i}_4 \neq {\bf i}_2 = {\bf i}_3$.
There can also be arbitrary 4-fermion terms with ${\bf i}_1 \neq {\bf i}_2 \neq {\bf i}_3 \neq {\bf i}_4$. 

We organize the calculation by exploiting the localization of the flat band WFs, which is controlled by 
$E_0$, the gap to bands outside the low-energy subspace, the same parameter that justified the single-particle projection to flat bands.
Exponentially localized WFs decay in space with with a characteristic length $\sqrt{\Omega}$, and the spread $\Omega$ decreases with increasing $E_0$. 
Thus the on-site value of the WF $|W({\bf 0})|$ is much larger than its value $ |W({\bf a})|$ even one lattice spacing away, 
with $ |W({\bf a})|/|W({\bf 0})| \sim \exp(- c E_0/t)$ (where $c$ is a constant). We present numerical evidence for this exponential decay 
with $E_0/t$ in Fig.~\ref{fig-lieb-ut-jnn}(c) for the Lieb lattice.

This behavior of the WFs makes renormalized Hubbard interaction
\begin{equation}
|\widetilde{U}| = |U| \sum_{{\bf i}^\prime \alpha} | W_{\ell\alpha}({\bf i}^\prime)|^4
\end{equation}
the largest amongst all interaction terms, since all other terms involve WFs centered on least two different sites.
All other interaction terms have at least two of the WFs in Eq.~\eqref{v1234-wf} evaluated in different unit cells and thus
the resulting interaction is exponentially smaller than $|\widetilde{U}|$. As an example, consider the pair-hopping term
\begin{equation}
 \mathcal{H}_{ph} = - \sum\limits_{ {\bf i}, {\bf j} } \mathcal{K}_{ij} c^{\dag}_{ {\bf i} \ell \uparrow } c^{\dag}_{ {\bf i} \ell \downarrow } 
 c^{\phantom\dag}_{ {\bf j} \ell \downarrow }  c^{\phantom\dag}_{ {\bf j} \ell \uparrow }
 \end{equation}
with the non-negative pair hopping amplitude
 \begin{equation}
 \mathcal{K}_{ij} = |U| \sum_{ {\bf i}^\prime \alpha} | W_{\ell\alpha}({\bf i}-{\bf i}^\prime)|^2 | W_{\ell\alpha}({\bf j}-{\bf i}^\prime)|^2. \label{def-kij}
  \end{equation}
 Due to the localization of the WFs, $ \mathcal{K}_{ij}$ for arbitrary ${\bf i}$ and ${\bf j}$ is 
 exponentially smaller than the nearest neighbor (NN) pair hopping $ \mathcal{K}_{NN}$, 
 corresponding to ${\bf i}$ and {\bf j} NN sites.
 However, even the dominant pair hopping $\mathcal{K}_{NN}\sim |U|\exp(- 2c E_0/t)$ is exponentially small
 compared to the ``on-site" Hubbard $|\widetilde{U}|$.
 In sec.~\ref{app-Lieb}, we provide quantitative evidence in support of this argument by 
 studying the ratio of $\widetilde{U}$ to other NN interactions as a function of gap $E_0$ for the Lieb lattice; see Fig.~\ref{fig-lieb-ut-jnn}.

The dominance of the  on-site attractive interaction, $\widetilde{U}$, forces the electrons to form tightly-bound pairs,
and the low energy Hilbert space consists of states with doubly occupied or empty sites, i.e., no single occupancy.
By analogy with the well-known ``lower Hubbard band" in the repulsive case, we
will call the ``no single occupancy" subspace as our lower Hubbard band in the attractive case.
Any four-fermion interaction term that involves a singly occupied site necessarily takes one outside this
low energy subspace, and can hence be ignored.

In addition to pair hopping, the other interaction terms that preserve the number of pairs are: density-density interaction, $\mathcal{H}_{\rho\rho}$ and spin-flip interaction $\mathcal{H}_{\uparrow\downarrow}$:
\begin{eqnarray}
    \mathcal{H}_{\rho\rho} &=& -|U| \sum\limits_{ {\bf i}, {\bf j} } \left( \sum\limits_{{\bf i}^\prime \alpha} |W_{\ell\alpha}({\bf i}-{\bf i}^\prime)|^2 |W_{\ell\alpha}({\bf j}-{\bf i}^\prime)|^2 \right)  c^\dag_{ {\bf i} \ell \uparrow } c^{\phantom\dag}_{ {\bf i} \ell \uparrow } c^\dag_{ {\bf j} \ell \downarrow } c^{\phantom\dag}_{ {\bf j} \ell \downarrow },\\
     \mathcal{H}_{\uparrow\downarrow} &=& -|U| \sum\limits_{ {\bf i}, {\bf j} } \left( \sum\limits_{ {\bf i}^\prime\alpha} W_{\ell\alpha}({\bf i}-{\bf i}^\prime)^2 (W_{\ell\alpha}({\bf j}-{\bf i}^\prime)^*)^2 \right) c^\dag_{ {\bf i} \ell \uparrow } c^{\phantom\dag}_{ {\bf i} \ell \downarrow } c^\dag_{ {\bf j} \ell \downarrow } c^{\phantom\dag}_{ {\bf j} \ell \uparrow }
    \end{eqnarray}
The arguments given above again imply that longer range interactions are exponentially smaller than the NN interactions, which themselves
are exponentially small compared to the on-site $|\widetilde{U}|$.

Hence, the low energy Hamiltonian is given by
\begin{eqnarray}
    \mathcal{H}_{\text{low}} &=& -|\widetilde{U}| \sum\limits_{ {\bf i} } c^{\dag}_{ {\bf i} \ell \uparrow } c^{\dag}_{ {\bf i} \ell \downarrow } c^{\phantom\dag}_{ {\bf i} \ell \downarrow } c^{\phantom\dag}_{ {\bf i} \ell \uparrow }   -  \sum\limits_{ {\bf i}, {\bf j} } \mathcal{K}_{ij} c^{\dag}_{ {\bf i} \ell \uparrow } c^{\dag}_{ {\bf i} \ell \downarrow } c^{\phantom\dag}_{ {\bf j} \ell \downarrow } c^{\phantom\dag}_{ {\bf j} \ell \uparrow } + \mathcal{H}_{\rho\rho} + \mathcal{H}_{\uparrow\downarrow}
\end{eqnarray}
where in the last three terms involving two different sites ${\bf i}$ and {\bf j}, only the NN terms need to be retained. 
The bandwidth for excitations in this low-energy subspace is set by $\mathcal{K}_{NN} \ll |U| \ll E_0$

The reason why we only show the pair hopping explicitly is that it is the only term that couples to the external vector potential as we show next.
Thus it plays the central role in computing optical spectral weight using 
\begin{equation}
\widetilde{D}_{\rm low} = {{\hbar^2}\over{2 \pi e^2}} \int_0^\Lambda d\omega~{\rm Re}~\sigma(\omega) 
= {{\hbar^2}\over{8e^2 N_c}} \left\langle {{\partial^2 \mathcal{H}_{\rm low}}\over{\partial A_\mu^2}} \right\rangle.
\label{optical-sp-wt-app}
\end{equation}

This is a good point to recap the derivation of our effective low-energy Hamiltonian $\mathcal{H}_{\text{low}}$ and to understand why it will
give us a tight bound on the low energy optical spectral weight. At the first step, described in 
sec.~\ref{app-projection}, we projected down to the subspace spanned by the flat band states $\{ \ell \uparrow, \ell \downarrow \}$.
This is controlled by the large band gap $E_0$ separating the partially-filled flat bands from empty and filled bands. 
In this section, we analyzed the various four-fermion terms in the effective Hamiltonian and organized them by exploiting the
 exponential localization of WFs, which is also controlled by large $E_0$.
This finally led to our effective low-energy Hamiltonian $\mathcal{H}_{\rm low}$ that acts within the lower Hubbard band (with no single occupancy).
Thus $\mathcal{H}_{\rm low}$ describes the physics on energy/temperature scales with an effective
cut off $\Lambda < |\widetilde{U}|$.  We will use $\mathcal{H}_{\rm low}$ to compute 
the intraband optical spectral weight in the lower Hubbard band with a (pair hopping) bandwidth $\mathcal{K}_{NN}$.
}

\section{Vector potential in lattice models}\label{app-vecPotLatModel}

In the presence of scalar potential $\varphi({\bf r},t)$ and a vector potential ${\bf A}({\bf r},t)$, the Hamiltonian is transformed:
\begin{equation}
\mathcal{H} \longrightarrow \mathcal{H}({\bf A},\varphi) =\sum\limits_{ {\bf i} {\bf j} \alpha\beta} t_{\alpha\beta}( {\bf i}- {\bf j}; {\bf A} ) d^\dag_{ {\bf i} \alpha} d^{\phantom\dag}_{ {\bf j} \beta} + \sum\limits_{ {\bf i}_j \alpha_j} V_{ 1234 }( {\bf i}_j, \alpha_j; {\bf A} ) d^\dag_{ {\bf i}_1 \alpha_1} d^{\phantom\dag}_{ {\bf i}_2 \alpha_2} d^\dag_{ {\bf i}_3 \alpha_3} d^{\phantom\dag}_{ {\bf i}_4 \alpha_4}+ e\sum\limits_{ {\bf i}\alpha } \varphi( {\bf r}_{ {\bf i}\alpha},t ) d^\dag_{ {\bf i} \alpha} d^{\phantom\dag}_{ {\bf i} \alpha}. \label{ham-gauge-1}
\end{equation}
Our goal is to use gauge invariance to determine how determine how ${\bf A}$ enters 
 $  t_{\alpha\beta}( {\bf i}- {\bf j}; {\bf A} )$ and $V_{ 1234 }( {\bf i}_j, \alpha_j; {\bf A} )$. 
As expected  $  t_{\alpha\beta}( {\bf i}- {\bf j}; {\bf A} )$ will acquire the Peierls' phase factor, 
but the ${\bf A}$-dependence of the interaction term is less well known and worth deriving carefully,
even though the final answer is obvious. The interaction term in the original Hamiltonian (Eq.~\eqref{Hfull-1-int}) 
will have no ${\bf A}$-dependence if we start with density-density or spin-spin interaction.
 Even in this case the interaction term in the projected low-energy Hamiltonian
Eq.~\eqref{H-low} will acquire non-trivial ${\bf A}$-dependence, and this is what we are really interested in.
The analysis presented here is general and applies to an arbitrary lattice Hamiltonians provided the orbitals $\{\alpha\}$ are exponentially localized in real space.

We are interested in linear response to electromagnetic fields in the ${\bf q} \to 0$ limit.
For the discussion, it suffices to consider the case of a spatially uniform electric field ${\bf E}$. Recall that a uniform ${\bf E}$ has two gauge choices:
\begin{enumerate}
\item The ``length gauge" $\varphi({\bf r}) \neq 0$ and ${\bf A}({\bf r},t) = {\bf 0}$, with ${\bf E} = - \boldsymbol\nabla \varphi$, where
\begin{equation}
\mathcal{H}({\bf A}=0,\varphi) =\sum\limits_{ {\bf i} {\bf j} \alpha\beta} t_{\alpha\beta}( {\bf i}- {\bf j} ) d^\dag_{ {\bf i} \alpha} d^{\phantom\dag}_{ {\bf j} \beta} + \sum\limits_{ {\bf i}_j \alpha_j} V_{ 1234 }( {\bf i}_j, \alpha_j ) d^\dag_{ {\bf i}_1 \alpha_1} d^{\phantom\dag}_{ {\bf i}_2 \alpha_2} d^\dag_{ {\bf i}_3 \alpha_3} d^{\phantom\dag}_{ {\bf i}_4 \alpha_4}+ e\sum\limits_{ {\bf i}\alpha } \varphi( {\bf r}_{ {\bf i}\alpha} ) d^\dag_{ {\bf i} \alpha} d^{\phantom\dag}_{ {\bf i} \alpha} \label{ham-gauge-L}
\end{equation}

\item The ``velocity gauge" $\varphi({\bf r}) = 0$ and ${\bf A}({\bf r},t) = {\bf A}(t)$, with ${\bf E} = - \partial_t {\bf A}$, where 
\begin{equation}
\mathcal{H}({\bf A},\varphi=0) =\sum\limits_{ {\bf i} {\bf j} \alpha\beta} t_{\alpha\beta}( {\bf i}- {\bf j}; {\bf A} ) d^\dag_{ {\bf i} \alpha} d^{\phantom\dag}_{ {\bf j} \beta} + \sum\limits_{ {\bf i}_j \alpha_j} V_{ 1234 }( {\bf i}_j, \alpha_j;{\bf A} ) d^\dag_{ {\bf i}_1 \alpha_1} d^{\phantom\dag}_{ {\bf i}_2 \alpha_2} d^\dag_{ {\bf i}_3 \alpha_3} d^{\phantom\dag}_{ {\bf i}_4 \alpha_4} \label{ham-gauge-V}
\end{equation}
\end{enumerate}

The Hamiltonian in the two gauges are simply related by $U(1)$ gauge transformation.
We can go from the velocity to length gauge by choosing $\chi({\bf r},t)$ such that 
$\partial_t \chi({\bf r}_{ {\bf i}\alpha},t) = \varphi({\bf r}_{ {\bf i}\alpha},t)$ and $\boldsymbol\nabla \chi({\bf r}_{ {\bf i}\alpha},t) = {\bf A}(t)$.
The field operators also change under this gauge transformation
\begin{equation}
d^{\phantom\dag}_{ {\bf i} \alpha} \longrightarrow \exp\left(i \frac{e}{\hbar} \chi( {\bf r}_{{\bf i}\alpha},t) \right) d^{\phantom\dag}_{ {\bf i} \alpha}
\end{equation}
which forces the hopping elements to pick up phase factors in order to make Eq.~ \ref{ham-gauge-L} and Eq.~\eqref{ham-gauge-V} equivalent.
For the kinetic energy this is the familiar Peierls phase factor with a uniform ${\bf A}$
\begin{eqnarray}
t_{\alpha\beta}( {\bf i}- {\bf j}; {\bf A} ) = t_{\alpha\beta}( {\bf i}- {\bf j}) e^{ -i \frac{e}{\hbar} \Big[\chi( {\bf r}_{{\bf i}\alpha},t) - \chi( {\bf r}_{{\bf j}\beta},t) \Big]} & = & t_{\alpha\beta}( {\bf i}- {\bf j}) \exp\left( -i \dfrac{e}{\hbar} \int\limits_{ {\bf r}_{ {\bf i} \alpha} }^{ {\bf r}_{ {\bf j}\beta}} d {\bf r} \cdot \boldsymbol\nabla \chi( {\bf r},t) \right) \\
&=& t_{\alpha\beta}( {\bf i}- {\bf j}) \exp\left[ -i \dfrac{e}{\hbar} ({\bf r}_{ {\bf j}\beta} - {\bf r}_{ {\bf i} \alpha}) \cdot {\bf A}(t) \right]. \label{vp1}
\end{eqnarray}
As we will see later, the orbital dependence in the phase factor is crucial in interpreting the ``minimal substitution'' for Bloch Hamiltonians.
The same argument also leads to the ${\bf A}$-dependence for the interaction term 
\begin{equation}
V_{ 1234 }( {\bf i}_j, \alpha_j;{\bf A} ) = V_{ 1234 }( {\bf i}_j, \alpha_j)
 \ \exp\left[ -i \dfrac{e}{\hbar}({\bf r}_{ {\bf i}_4 \alpha_4} - {\bf r}_{ {\bf i}_3 \alpha_3} + {\bf r}_{ {\bf i}_2 \alpha_2} - {\bf r}_{ {\bf i}_1 \alpha_1})  \cdot {\bf A}(t) \right]. \label{vecpot-int}
\end{equation}
The result is easy to understand intuitively for an interaction term of the form $V_{ 1234 }( {\bf i}_j, \alpha_j ) d^\dag_{ {\bf i}_1 \alpha_1} d^{\phantom\dag}_{ {\bf i}_2 \alpha_2} d^\dag_{ {\bf i}_3 \alpha_3} d^{\phantom\dag}_{ {\bf i}_4 \alpha_4}$.
We note that density-density or spin-spin interactions have 
${\bf r}_{ {\bf i}_1 \alpha_1} = {\bf r}_{ {\bf i}_2 \alpha_2} $ and ${\bf r}_{ {\bf i}_3 \alpha_3} = {\bf r}_{ {\bf i}_4 \alpha_4} $,
which makes them independent of the vector potential. 
On the other hand, correlated hops or pair hopping terms pick up an ${\bf A}$-dependence.
Pair hopping terms, with ${\bf r}_{ {\bf i}_1 \alpha_1} = {\bf r}_{ {\bf i}_3 \alpha_3} $, ${\bf r}_{ {\bf i}_2 \alpha_2} = {\bf r}_{ {\bf i}_4 \alpha_4} $, have twice the phase as that of a single particle hop.

\section{Low-energy Spectral Weight}\label{app-proj-D}

\rev{
The results of sec.~\ref{app-vecPotLatModel} show that the {\em only}  term in the low energy Hamiltonian that
couples to the vector potential ${\bf A}$ is the pair hopping term $\mathcal{K}_{ij}$. 
The density-density and spin-flip interactions do not couple to ${\bf A}$ and do not 
impact ${{\partial^2 \mathcal{H}_{\rm low}}/{\partial A_\mu^2}}$. This is why we did not even display these terms 
explicitly in the main text.

 Using Eq.~\eqref{vecpot-int}, we find
\begin{equation}
 \mathcal{H}_{\text{low}} \longrightarrow \mathcal{H}_{\text{low}}({\bf A}) = -|\widetilde{U}| \sum\limits_{ {\bf i} } c^{\dag}_{ {\bf i} \ell \uparrow } c^{\dag}_{ {\bf i} \ell \downarrow } c^{\phantom\dag}_{ {\bf i} \ell \downarrow } c^{\phantom\dag}_{ {\bf i} \ell \uparrow }   -  \sum\limits_{ {\bf i}, {\bf j} } e^{ i2e {\bf A}\cdot \left( {\bf i}-{\bf j} \right) } \mathcal{K}_{ij} c^{\dag}_{ {\bf i} \ell \uparrow } c^{\dag}_{ {\bf i} \ell \downarrow } c^{\phantom\dag}_{ {\bf j} \ell \downarrow } c^{\phantom\dag}_{ {\bf j} \ell \uparrow } + \mathcal{H}_{\rho\rho} + \mathcal{H}_{\uparrow\downarrow}.
\end{equation}
Next, we use Eq.~\eqref{optical-sp-wt-app} to find}
\begin{eqnarray}
[\boldsymbol{\widetilde{D}}_\text{low}]_{\mu\nu} &=& \dfrac{1}{N_c} \sum\limits_{ {\bf i}, {\bf j} } \left( {\bf i}-{\bf j} \right)_\mu \left( {\bf i}-{\bf j} \right)_\nu \mathcal{K}_{ij} \left\langle c^{\dag}_{ {\bf i} \ell \uparrow } c^{\dag}_{ {\bf i} \ell \downarrow } c^{\phantom\dag}_{ {\bf j} \ell \downarrow } c^{\phantom\dag}_{ {\bf j} \ell \uparrow } \right\rangle \label{proj-D-1}
\end{eqnarray}
where $\mu,\nu$ are spatial indices. The factor of $1/4$ in front is cancelled by $(2e)^2$ coming from charge of the pair.
The spectral weight is, in general, a tensor whose trace is important to us.
Since we will be focussing on 2D in this paper, we define
\begin{equation}
\widetilde{D}_\text{low} = \dfrac{1}{2} \text{Tr}[\boldsymbol{\widetilde{D}}_\text{low}].\label{def-dtilde-trace}
\end{equation}
This idea can easily be extended to dimensions other than 2 by a simple pre-factor. 
$\widetilde{D}_\text{low}$ should be interpreted as a sum rule for the trace of the dynamical conductivity tensor.
Since the integral of a dissipative response is positive by second law of thermodynamics, we can use triangle inequality to derive an upper bound
\begin{eqnarray}
\widetilde{D}_\text{low} &\leq& \dfrac{1}{2 N_c}\sum\limits_{ {\bf i}, {\bf j} } \left( {\bf i}-{\bf j} \right)^2 \mathcal{K}_{ij} \left| \left\langle c^{\dag}_{ {\bf i} \ell \uparrow } c^{\dag}_{ {\bf i} \ell \downarrow } c^{\phantom\dag}_{ {\bf j} \ell \downarrow } c^{\phantom\dag}_{ {\bf j} \ell \uparrow } \right\rangle \right|.
\end{eqnarray}.

Now we will plug in $\mathcal{K}_{ij}$ that has an implicit sum on ${\bf i}^\prime$ (see Eq.~\eqref{def-kij}). After re-labelling the sum ${\bf i} \longrightarrow {\bf i} + {\bf i}^\prime$, ${\bf j} \longrightarrow {\bf j} + {\bf i}^\prime$, we will get
\begin{eqnarray}
\widetilde{D}_\text{low} &\leq & \dfrac{|U|}{2} \sum\limits_{ {\bf i}, {\bf j} } \left( {\bf i}-{\bf j} \right)^2 \left( \sum_{ \alpha} | W_{\ell\alpha}({\bf i})|^2 | W_{\ell\alpha}({\bf j})|^2 \right) \left( \dfrac{1}{N_c} \sum\limits_{ {\bf i}^\prime } \left|\left\langle c^{\dag}_{ {\bf i}+{\bf i}^\prime \ell \uparrow } c^{\dag}_{ {\bf i}+{\bf i}^\prime \ell \downarrow } c^{\phantom\dag}_{ {\bf j}+{\bf i}^\prime \ell \downarrow } c^{\phantom\dag}_{ {\bf j}+{\bf i}^\prime \ell \uparrow } \right\rangle \right| \right) \\
&=& \dfrac{|U|}{2} \sum\limits_{ {\bf i}, {\bf j} } \left( {\bf i}-{\bf j} \right)^2 \left( \sum_{ \alpha} | W_{\ell\alpha}({\bf i})|^2 | W_{\ell\alpha}({\bf j})|^2 \right) \left|\left\langle c^{\dag}_{ {\bf i} \ell \uparrow } c^{\dag}_{ {\bf i} \ell \downarrow } c^{\phantom\dag}_{ {\bf j} \ell \downarrow } c^{\phantom\dag}_{ {\bf j} \ell \uparrow } \right\rangle \right|.
\end{eqnarray}

Next, we continue along the same direction and bound the expectation value from above.
Notice that for any two operators $A$ and $B$, the thermal expectation value can be seen as an inner product
\begin{equation}
(A,B) = \left\langle A^\dag B \right\rangle = \text{Tr}\left[ e^{ -\beta \mathcal{H} } A^\dag B \right].
\end{equation}
This allows us to use Cauchy-Schwarz inequality to derive
\begin{equation}
\left| \left\langle A^\dag B \right\rangle \right| \leq  \sqrt{ \left\langle A^\dag A \right\rangle \left\langle B^\dag B \right\rangle   } \label{CS-inequality}.
\end{equation}
We can go one step further if the operators are themselves quadratic, that is, $A = c^{\phantom\dag}_2  c^{\phantom\dag}_1 $ and $B = c^{\phantom\dag}_3 c^{\phantom\dag}_4$, where $1,2,3,4$ are labels such that $1 \neq 2 \neq 3 \neq 4$. 
\begin{eqnarray}
\left|\left\langle c^\dag_1 c^\dag_2 c^{\phantom\dag}_3 c^{\phantom\dag}_4 \right\rangle \right| &\leq & \sqrt{ \left\langle c^\dag_1 c^\dag_2 c^{\phantom\dag}_2 c^{\phantom\dag}_1 \right\rangle \left\langle c^\dag_4 c^\dag_3 c^{\phantom\dag}_3 c^{\phantom\dag}_4 \right\rangle  } = \sqrt{ \left\langle c^\dag_1 c^{\phantom\dag}_1 c^\dag_2 c^{\phantom\dag}_2  \right\rangle \left\langle c^\dag_3 c^{\phantom\dag}_3 c^\dag_4 c^{\phantom\dag}_4 \right\rangle  }  \\
&=& \sqrt[4]{ \left \langle c^\dag_1 c^{\phantom\dag}_1 c^\dag_1 c^{\phantom\dag}_1 \right\rangle \left \langle c^\dag_2 c^{\phantom\dag}_2 c^\dag_2 c^{\phantom\dag}_2  \right\rangle \left \langle c^\dag_3 c^{\phantom\dag}_3 c^\dag_3 c^{\phantom\dag}_3 \right\rangle \left \langle c^\dag_4 c^{\phantom\dag}_4 c^\dag_4 c^{\phantom\dag}_4 \right\rangle   } = \sqrt[4]{n_1 n_2 n_3 n_4}
\end{eqnarray}
where $n_i = \left\langle c^\dag_i c^{\phantom\dag}_i \right\rangle$ for $i=1,\cdots,4$ and we have used $\hat{n}_i^2 = \hat{n}_i$ to derive the last equality.
Applying this result to Eq.~\eqref{proj-D-1}, we get
\begin{equation}
\widetilde{D}_\text{low} \leq \dfrac{|U|}{2} \sum\limits_{ {\bf i}, {\bf j} } \left( {\bf i}-{\bf j} \right)^2 \left( \sum_{ \alpha} | W_{\ell\alpha}({\bf i})|^2 | W_{\ell\alpha}({\bf j})|^2 \right)  \sqrt[4]{ n_{ {\bf i} \ell \uparrow } n_{ {\bf i} \ell \downarrow } n_{ {\bf j} \ell \downarrow } n_{ {\bf j} \ell \uparrow } } 
\end{equation}
where we can use 
\begin{equation}
\sqrt[4]{ n_{ {\bf i} \ell \uparrow } n_{ {\bf i} \ell \downarrow } n_{ {\bf j} \ell \downarrow } n_{ {\bf j} \ell \uparrow } } \leq \dfrac{1}{4} \left( n_{ {\bf i} \ell \uparrow } + n_{ {\bf i} \ell \downarrow } + n_{ {\bf j} \ell \downarrow } + n_{ {\bf j} \ell \uparrow } \right) = \dfrac{n}{2} \label{n-amgm}.
\end{equation}
\rev{The last equality follows from the combination of translation invariance of density expectation values ($n_{ {\bf i} \ell \sigma } = n_{\sigma}$) and time-reversal $n_\uparrow = n_\downarrow = n/2$, resulting in}
\begin{equation}
\widetilde{D}_\text{low} \leq  |U| \dfrac{n}{4} \left( \sum\limits_{ {\bf i}, {\bf j},\alpha } \left( {\bf i}-{\bf j} \right)^2| W_{\ell\alpha}({\bf i})|^2 | W_{\ell\alpha}({\bf j})|^2  \right) = |U| \dfrac{n}{4} \mathcal{D} \label{proj-D-2}.
\end{equation}
where $n$ is the density of electrons in the flat band.
The final piece in the puzzle is to identify the term inside the parenthesis, $\mathcal{D}$, as a part of the spread functional in Eq.~\eqref{def-Omega}.
\begin{eqnarray}
\Omega = \dfrac{1}{2}\sum\limits_{ {\bf i}, \alpha } \sum\limits_{ {\bf j}, \beta} ( {\bf i} - {\bf j} )^2 |W_{ \ell \alpha }( {\bf i} )|^2 |W_{ \ell \beta }( {\bf j} )|^2  \geq  \dfrac{1}{2} \sum\limits_{ {\bf i},{\bf j}, \alpha} ({\bf i} - {\bf j})^2 |W_{ \ell \alpha }( {\bf i} )|^2 |W_{ \ell \alpha }( {\bf j} )|^2 = \dfrac{\mathcal{D}}{2} \label{s2:Omega-O}.
\end{eqnarray}
that gives the final answer
\begin{equation}
\widetilde{D}_\text{low} \leq \dfrac{ n }{2} |U| \Omega. \label{proj-spt-bound}
\end{equation}
\rev{
We can further use a particle-hole transformation, $c_i \longrightarrow h_i^\dag$ to derive tighter bounds:
\begin{eqnarray}
\left|\left\langle c^\dag_{ {\bf i} \ell \uparrow } c^\dag_{{\bf i} \ell \downarrow} c^{\phantom\dag}_{{\bf j} \ell \downarrow} c^{\phantom\dag}_{{\bf j} \ell \uparrow} \right\rangle \right| =\left|\left\langle h^\dag_{ {\bf i} \ell \uparrow } h^\dag_{{\bf i} \ell \downarrow} h^{\phantom\dag}_{{\bf j} \ell \downarrow} h^{\phantom\dag}_{{\bf j} \ell \uparrow} \right\rangle \right|&\leq & \sqrt[4]{ (1-n_{{\bf i} \ell \uparrow} ) (1-n_{{\bf i} \ell \downarrow}) (1-n_{{\bf j} \ell \downarrow}) (1-n_{{\bf j} \ell \uparrow}) } \\
&\leq & \dfrac{1}{4} \left( 4 - (n_{{\bf i} \ell \uparrow}+n_{{\bf i} \ell \downarrow} + n_{{\bf j} \ell \downarrow} + n_{{\bf j} \ell \uparrow})  \right) = 1 - \dfrac{n}{2}.
\end{eqnarray}
The inequality greatly improves our bound for systems with density $n>1$,
\begin{equation}
\widetilde{D}_\text{low} \leq \dfrac{ \widetilde{n} }{2} |U| \Omega, \quad \text{where } \widetilde{n} = \text{min}(n,2-n). 
\end{equation}
}

\rev{
Finally, we briefly remark on how our results can be generalized to the more general attractive Hamiltonian 
\begin{equation}
\mathcal{H}_\text{int} = - |U| \sum\limits_{ {\bf i} \alpha } f_\alpha\hat{n}_{ {\bf i}\alpha \uparrow } \hat{n}_{ {\bf i} \alpha \downarrow}
\label{H-general}
\end{equation}
with orbitally-dependent attraction: $0 \leq f_\alpha \leq 1$.
The usual attractive Hubbard model that we have focused on in the rest of the paper 
corresponds to $f_\alpha \equiv 1$ for all orbitals $\alpha$.

Let us look at the case with at least one non-zero $f_\alpha$
for an $\alpha$ where the flat-band has non-zero support i.e., $|W_{\ell\alpha}({\bf 0})| > 0$.
The renormalized Hubbard interaction is then given by $|\widetilde{U}| = |U| \sum_{{\bf i}^\prime \alpha} f_\alpha | W_{\ell\alpha}({\bf i}^\prime)|^4 $ and 
pair hopping integrals $\mathcal{K}_{ij} = |U| \sum_{{\bf i}^\prime \alpha} f_\alpha | W_{\ell\alpha}({\bf i}-{\bf i}^\prime)|^2 | W_{\ell\alpha}({\bf j}-{\bf i}^\prime)|^2$.
All the arguments of this section then go through, and we use the inequality $f_\alpha \leq 1$ to derive:
\begin{equation}
\widetilde{D}_\text{low} \leq |U| \dfrac{\widetilde{n}}{4} \sum\limits_{ {\bf i}, {\bf j} } \left( {\bf i}-{\bf j} \right)^2 \left( \sum\limits_\alpha f_\alpha |W_{\ell\alpha}({\bf i})|^2 | W_{\ell\alpha}({\bf j})|^2 \right) \leq  |U| \dfrac{\widetilde{n}}{4} \sum\limits_{ {\bf i}, {\bf j} } \left( {\bf i}-{\bf j} \right)^2 \left( \sum\limits_\alpha |W_{\ell\alpha}({\bf i})|^2 | W_{\ell\alpha}({\bf j})|^2 \right) \leq \dfrac{\widetilde{n}}{2} |U| \Omega.
\end{equation}
}

\section{Lieb Lattice}\label{app-Lieb}

We will now illustrate the formalism with the Lieb lattice.
It is an unbalanced bipartite lattice with the Bloch Hamiltonian
\begin{equation}
    \mathcal{H}_K = \sum\limits_{ {\bf k} \sigma} \Psi_{ {\bf k} }^\dag t({\bf k}) \Psi_{ {\bf k} }^{\phantom\dag}, \quad t({\bf k}) = \begin{pmatrix}
    0 & f^*(k_x) & f(k_y) \\
    f(k_x) & 0 & 0 \\
    f^*(k_y) & 0 & 0 
    \end{pmatrix} \label{Lieb-full-K}.
\end{equation}
Here $A,B,C$ are the three sublattices, $\Psi_{ {\bf k} } = \left( d_{ {\bf k} A }, d_{ {\bf k} B }, d_{ {\bf k} C } \right)^T$,  and $f(x) = t[(1+\delta) + (1- \delta) e^{ i x } ]$.
The single-particle dispersion (see Fig.~1 of main text) has an exactly flat band separated from other bands by a gap $E_0 = 2\sqrt{2}\delta t$ that can be controlled by staggering $\delta$. The flat band wavefunction is given by
\begin{equation}
| \ell {\bf k} \rangle = \dfrac{1}{\sqrt{|f(k_x)|^2 + |f(k_y)|^2}} \begin{pmatrix}
0 \\ f^*(k_y) \\ -f(k_x)
\end{pmatrix}. \label{lief-flat-band-waveF}
\end{equation}

\begin{figure}
\centering
\includegraphics[scale=1]{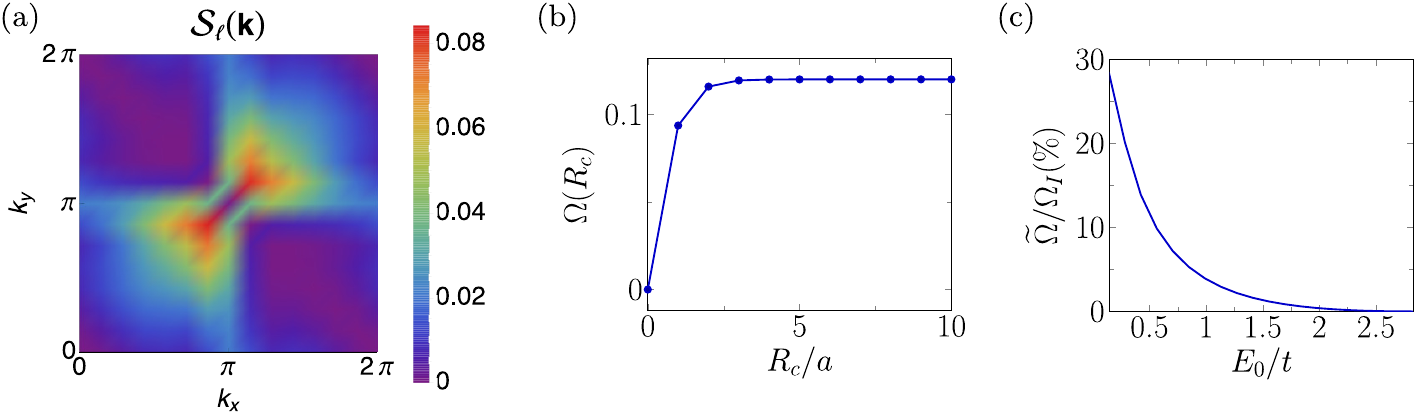}
\caption{
(a) Smoothness function $\mathcal{S}_\ell ({\bf k})$ defined in Eq.~\eqref{def-smoothness} for the flat-band wavefunction of 
Eq.~\eqref{lief-flat-band-waveF}.  (b). The Wannier functions are exponentially localized and result in a MV spread $\Omega(R_c)$ that converges 
rapidly as a function of the cutoff radius $R_c$; see Eq.~\eqref{omega-Rc}. (c). 
Variation of $\Omega- \Omega_I = \widetilde{\Omega}$ (see Eq.~(\ref{def-Omega-I})-(\ref{omega-quantum})) as a function of gap $E_0$. 
Here $\Omega_I$ is the trace of the quantum metric; see Eq.~\eqref{def-Omega-I}.
Note that the spread $\Omega$ approaches $\Omega_I$ as $E_0$ is increased.
}
\label{fig-lieb-wf}
\end{figure}

\begin{figure}
\centering
\includegraphics[scale=1]{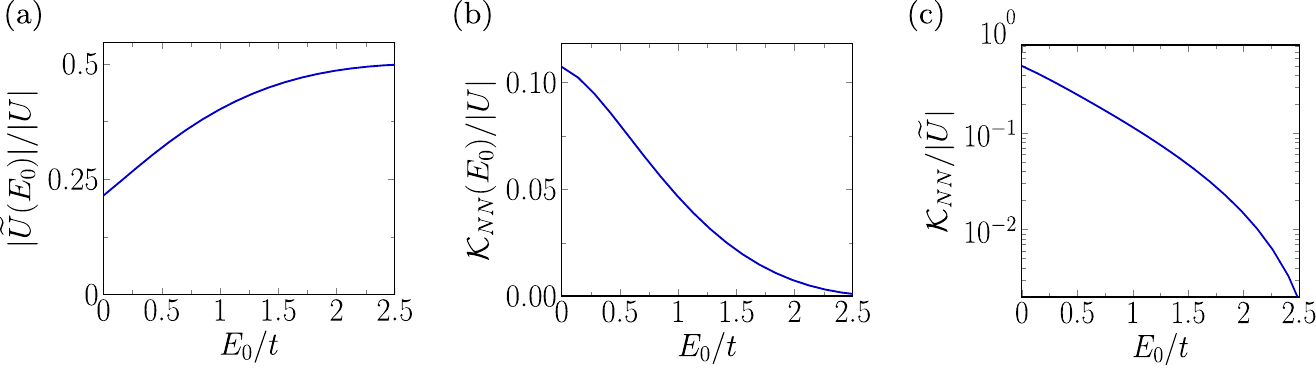}
\caption{
(a). The renormalized Hubbard interaction $|\widetilde{U}|$, the largest energy scale in the low-energy Hamiltonian, 
increases with increasing gap $E_0$. (b). \rev{
The NN pair hopping term $\mathcal{K}_{NN}$ decreases with $E_0$ and ultimately vanishes when $\delta = 1$.
(c). Log-linear plot showing the rapid decrease of $\mathcal{K}_{NN}/|\widetilde{U}|$ with $E_0$.
}
}
\label{fig-lieb-ut-jnn}
\end{figure}

As seen in Fig.~\ref{fig-lieb-wf}, the analytic expression is quite successful in providing exponentially localized Wannier functions. 
The spread $\Omega$ at $E_0 = t$ is within $5\%$ of the minimal spread $\Omega_I$.
Next, we can find the bound on spectral weight
\begin{equation}
\widetilde{D}_\text{low} \leq \widetilde{n} |U| \dfrac{\Omega}{2}. \label{swb-Lieb}
\end{equation}
The bound can be improved further. We introduce the orbital polarized $\langle {\bf r} \rangle_{ \ell \alpha} = \sum_{{\bf i}} {\bf i} | W_{\ell\alpha}({\bf i})|^2$ to re-write the spread functional as $\Omega = \langle {\bf r}^2 \rangle_\ell - (\sum_\alpha \langle {\bf r} \rangle_{ \ell \alpha}) \cdot(\sum_\beta \langle {\bf r} \rangle_{ \ell \beta})$ where $\alpha,\beta$ run over sublattices A,B and C.
Next, we use the mirror symmetry about the diagonal direction
to argue
\begin{equation}
\sum\limits_{\bf i} | W_{\ell B} ({\bf i}) |^2 = \sum\limits_{\bf i} | W_{\ell C} ({\bf i}) |^2 = \dfrac{1}{2}
\end{equation}
along with $W_{\ell A}({\bf i}) = 0 \; \forall {\bf i}$. We use these to write $\mathcal{D}$ as
\begin{eqnarray}
\mathcal{D} = \sum\limits_{ {\bf i}, \alpha} {\bf i}^2 |W_{\ell \alpha}({\bf i})|^2 - 2 \sum\limits_\alpha \left(\sum\limits_{ {\bf i}} {\bf i} |W_{\ell \alpha}({\bf i})|^2\right) \left(\sum\limits_{ {\bf j}} {\bf j} |W_{\ell \alpha}({\bf j})|^2\right)  = \langle {\bf r}^2 \rangle_\ell - 2\sum_\alpha |\langle {\bf r} \rangle_{ \ell \alpha}|^2 \label{d-omega-vareps}
\end{eqnarray}
which gives $\Omega = \mathcal{D} + \varepsilon$ where $\varepsilon = \left( \langle {\bf r} \rangle_{ \ell B} - \langle {\bf r} \rangle_{ \ell C} \right)^2 \geq 0$. The inequality $\mathcal{D} < \Omega$ improves our bound in Eq.~\eqref{swb-Lieb} by a factor of 2.
It turns out that $\varepsilon$ is quite small and $\mathcal{D}$ is within $0.01\%$ of $\Omega$.
In fact, we can make $\varepsilon =0$ by appropriately shifting the Wannier centers via a gauge transformation.
Therefore, we conclude that $\mathcal{D} = \Omega$ for Lieb lattice.

\rev{
We end this section by presenting an exact $T=0$ lower bound
involving the quantum metric for all fillings of the flat band $0<n<2$.
The lower bound uses the generalization of Lieb's theorem~\cite{Tasaki2020} which shows that the ground state of the
attractive Hubbard model on unbalanced bipartite lattices, like the Lieb lattice, exhibits off-diagonal long range
order (ORLRO)~\cite{ShenPRL1993}.
This can be seen from Theorem 10.8 in Tasaki's book~\cite{Tasaki2020} which states that in the thermodynamic limit at $T=0$
\begin{equation}
\dfrac{1}{|\Lambda|^2} \left\langle \left( \sum\limits_{ x \in \Lambda } c_{ x \downarrow } c_{ x \uparrow } \right)^\dag 
\left( \sum\limits_{ x^\prime \in \Lambda } c_{ x^\prime \downarrow } c_{ x^\prime \uparrow } \right) \right\rangle \geq \left( \dfrac{1+a}{2} - \nu \right) \left( \nu - \dfrac{1-a}{2} \right).
\label{tasaki-bound}
\end{equation}
We have used here the notation of ref.~\cite{Tasaki2020}  where
$\Lambda$ is the ``volume" of the system and $a$ is a dimensionless measure of the 
imbalance in the lattice, which for Lieb lattice is $a = 1/3$. 
The filling fraction $\nu$, measured from the bottom of the lower band, is given by $\nu = (2 + n)/6$ in terms of flat-band density $n$.
Thus the RHS of Eq.~\eqref{tasaki-bound} is $n(2-n)/{36}$.

ODLRO is defined via $P({\bf i}-{\bf j}) = \left\langle c^\dag_{ {\bf i} \uparrow} c^\dag_{ {\bf i} \downarrow} c^{\phantom\dag}_{ {\bf j} \downarrow} c^{\phantom\dag}_{ {\bf j} \uparrow}  \right\rangle \rightarrow n_{B0}$ as $|{\bf i}-{\bf j}| \rightarrow \infty$.
It is easy to see that, in the $|\Lambda| \rightarrow \infty$ limit, the correlation function on the LHS of Eq.~\eqref{tasaki-bound} is simply equal to the
$n_{B0}$. If we further assume that $P({\bf r}) \geq n_{B0} \;\forall {\bf r}$, we get
\begin{equation}
P({\bf i}-{\bf j}) = \left\langle c^\dag_{ {\bf i} \uparrow} c^\dag_{ {\bf i} \downarrow} c^{\phantom\dag}_{ {\bf j} \downarrow} c^{\phantom\dag}_{ {\bf j} \uparrow}  \right\rangle \geq n_{B0} \geq \dfrac{n(2-n)}{36}.
\end{equation}
Using this in Eq.~\eqref{proj-D-1}, together with the fact that $\mathcal{D} = \Omega$ (see Eq.~\eqref{d-omega-vareps}), we obtain obtain the lower bound
\begin{eqnarray}
\widetilde{D}_\text{low} &=&  \dfrac{|U|}{2} \sum\limits_{ {\bf i}, {\bf j},\alpha } \left( {\bf i}-{\bf j} \right)^2 | W_{\ell\alpha}({\bf i})|^2 | W_{\ell\alpha}({\bf j})|^2 \left\langle c^{\dag}_{ {\bf i} \ell \uparrow } c^{\dag}_{ {\bf i} \ell \downarrow } c^{\phantom\dag}_{ {\bf j} \ell \downarrow } c^{\phantom\dag}_{ {\bf j} \ell \uparrow } \right\rangle  \geq \dfrac{|U|}{2} \mathcal{D} \dfrac{n(2-n)}{36} = \dfrac{n(2-n)}{72} |U| \Omega\\
&\geq &  \dfrac{n(2-n)}{72} |U| \left(\dfrac{1}{N_c} \sum\limits_{ {\bf k}} \text{Tr}[ g_\ell ({\bf k})]\right)
\end{eqnarray}
where we have used Eq.~\eqref{omega-quantum} to lower bound the spread functional by the quantum metric.
}

\section{Projected Spectral Weight in Topological bands: $\pi$-flux model}\label{app-proj-weight-topo}

We now consider a time-reversal invariant model where the low-energy bands are almost flat and topological. 
When these bands are filled, the system describes a $\mathbb{Z}_2$ topological insulator, with an opposite Chern number of $\pm 1$ on the two spin bands.
We consider the regime where the narrow bands are partially filled.
It may seem that the non-trivial topology of the bands may pose an obstruction to exponentially localized Wannier functions.
However, we can circumvent the obstruction and find an exact bound on the superfluid stiffness as before.

The $\pi$-flux model was first introduced as an analog of the Haldane model on a square lattice, whose bands could be made flat using only a few longer-range hoppings \cite{NeupertPRL2011}.
It is a two-orbital model (see Fig.\ 2 of main text) where one sublattice sits at the plaquette center.
NN hopping $t_1$ induces a winding and NNN hopping $t_2$ provides a ${\bf k}$-dependent mass term. 
With these two, the Bloch Hamiltonian is given by ${\bf d}({\bf k})\cdot \boldsymbol\sigma$ where 
\begin{eqnarray}
d_x + i d_y &=& -t_1 \left[ e^{-i \phi } \left(e^{-i {\bf k} \cdot \bf a_1} +e^{-i {\bf k}\cdot {\bf a}_2}\right) + e^{i \phi } \left( 1 + e^{-i {\bf k}.({\bf a}_1 + {\bf a}_2)} \right) \right] \\
d_z &=& - 2 t_2 \left[ \cos( {\bf k}\cdot {\bf a}_1 ) -  \cos( {\bf k}\cdot {\bf a}_2 ) \right].
\end{eqnarray}
The phase $\phi$ induces a flux $\Phi = \pm 4 \phi$ in the red/blue palquettes shown in Fig.~2 of main text.
The topology of the bands can be seen from the ${\bf d}$ vector: $d_x + i d_y$ vanishes at points $X = (\pi,0)$ and $Y = (0,\pi)$, which carry opposite signs of the mass term $d_z$.
This results in a non-zero Chern number as long as $t_2 \neq 0$ and $\Phi \neq 2\pi$.
Of particular interest to us is the choice $t_2 = t_1/\sqrt{2}$, $\phi = \pi/4$ that corresponds to flux $\Phi = \pi$ through the plaquette (hence the name $\pi$-flux).
While the topology remains intact, the energy becomes
\begin{equation}
\epsilon_{\pm}({\bf k}) = \pm| {\bf d} ({\bf k}) |, \quad \text{where } | {\bf d} ({\bf k}) | = \sqrt{ \cos( 2{\bf k}\cdot {\bf a}_1) + \cos( 2 {\bf k}\cdot {\bf a}_2) + 6 }.
\end{equation}

These ${\bf k}\cdot 2{\bf a}_i$ harmonics can be induced from an intra-orbital fifth neighbor hopping
\begin{equation}
d_0 = -2 t_5 \left[ \cos( {\bf k}\cdot (2{\bf a}_1) ) +  \cos( {\bf k}\cdot(2 {\bf a}_2) ) \right].
\end{equation}
We can chose the parameter $t_5$ to almost cancel the dispersion of one of the bands.
$t_5 = (1-\sqrt{2})/4$ is the fine-tuned value that optimizes the flatness with bandwidth $w = 0.036 t_1$ and isolates the band with gap $E_0 = 3.94 t_1$ \cite{HofmannPRB2020}.
We will work in a regime where $w \ll |U| \ll E_0$ so that finite $w$ effects can essentially be ignored.

\begin{figure}
\centering
\includegraphics[scale=1]{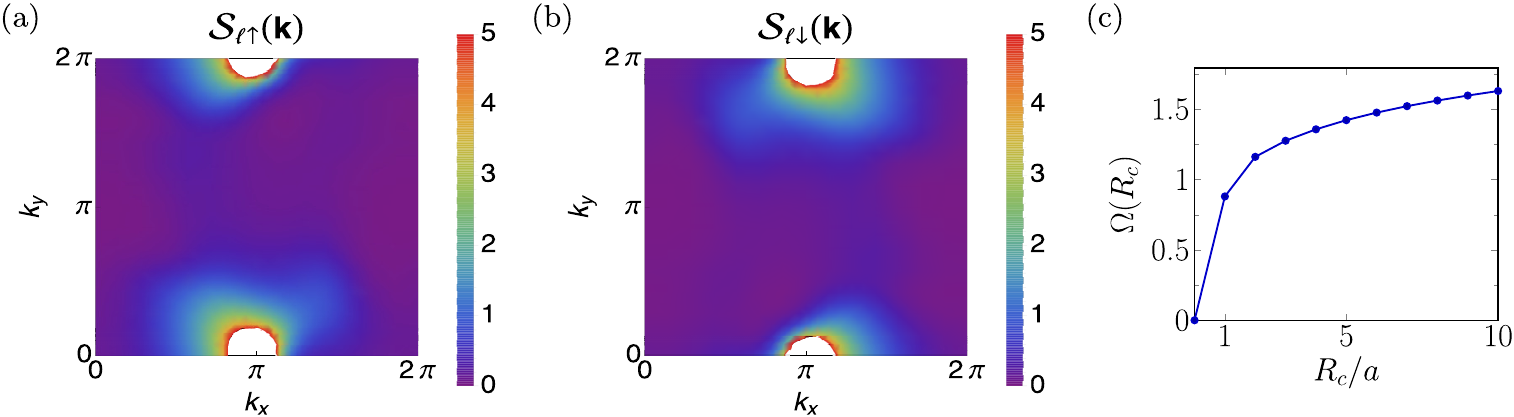}
\caption{(a)-(b) Smoothness function $\mathcal{S}_\ell ({\bf k})$ for the spin-polarized bands $| \ell \uparrow {\bf k} \rangle$ and 
$|\ell \downarrow {\bf k} \rangle$ of the $\pi$-flux model. We have chosen the gauge to pin the singularity at $X=(\pi,0)$. 
(c). The singularity results in Wannier functions with an algebraic decay, whose MV spread diverges, as can be seen
in the way that $\Omega(R_c)$ keeps on growing with as we increase the cut-off $R_c$. }
\label{fig-pf-updown}
\end{figure}

Spins carry opposite windings with $\phi_\uparrow = -\phi_\downarrow = \pi/4$ so that overall the system is time-reversal symmetric.
Since $\hat{S}_z$ is conserved, the $\mathbb{Z}_2$ invariant is $\nu = 1/2(\mathcal{C}_\uparrow - \mathcal{C}_\downarrow)$ for the lower band subspace.
The $\mathbb{Z}_2$-odd topology of the low-energy Hilbert space obstructs exponentially localized Wannier functions that are spin-polarized.
The obstruction manifests in a singularity in smoothness function $\mathcal{S}_{\ell\sigma}({\bf k})$ (see Fig.~\ref{fig-pf-updown}) for spin-polarized bands.
Different choices of the gauge can only move it to different points but cannot annihilate it.
The Wannier functions thus obtained are not localized.

\begin{figure}
\centering
\includegraphics[scale=1]{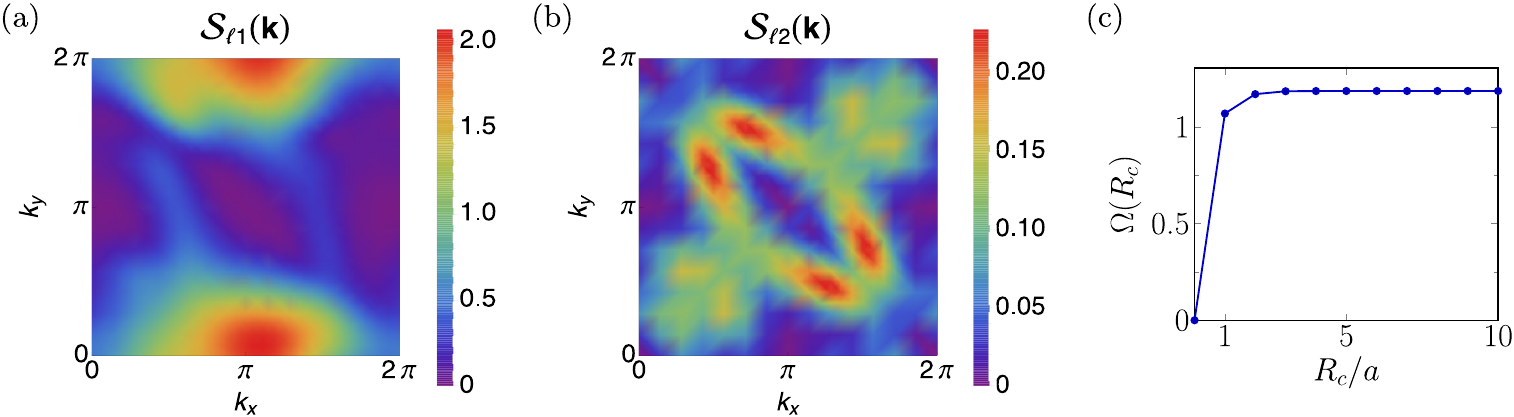}
\caption{(a)-(b) Smoothness function $\mathcal{S}_\ell ({\bf k})$ for the spin-mixed bands $| \ell 1 {\bf k} \rangle$ and $|\ell 2 {\bf k} \rangle$,
which are chosen so that there is no singular behavior. (c). We see that, in contrast to the case of the spin-eigenstates, the corresponding 
WFs are exponentially localized with a spread $\Omega(R_c)$ that converges rapidly with increasing $R_c$ to limiting value $\Omega = 1.19$. 
The minimum spread, or the quantum metric, is $\Omega_I = 1.03$. The fast convergence can be attributed to the large energy gap 
$E_0 = 3.94t_1$. }
\label{fig-pf-smooth}
\end{figure}

We employ the method outlined in ref \cite{SoluyanovPRB2012} to ``unwind'' the topology and find Wannier functions that are exponentially localized.
First, we fix a gauge so that both spin bands have the singularity at the $X$ point in the BZ.
Then we construct a $2\times 2$ unitary matrix $\mathcal{U}^\dag({\bf k}) = \{ | \ell \uparrow {\bf k} \rangle, | u \uparrow {\bf k} \rangle$ whose columns are the eigenvectors corresponding to the lower and upper bands of one spin sector. 
We fix the gauge of $| u \uparrow {\bf k} \rangle$ so that $\mathcal{S}_{u\uparrow}({\bf k})$ is also singular at $X$.
The matrix $\mathcal{U}^\dag({\bf k})$ is obviously topologically non-trivial and breaks time-reversal.
Finally, we use its inverse to ``rotate'' the spin bands and construct 
\begin{equation}
| \ell 1 {\bf k} \rangle = \mathcal{U}_{11}({\bf k}) | \ell \uparrow {\bf k} \rangle + \mathcal{U}_{21}({\bf k}) | \ell \downarrow {\bf k} \rangle, \quad | \ell 2 {\bf k} \rangle = \mathcal{U}_{12}({\bf k}) | \ell \uparrow {\bf k} \rangle + \mathcal{U}_{22}({\bf k}) | \ell \downarrow {\bf k} \rangle. \label{pf-rotate}
\end{equation}
The resulting spin-mixed ``rotated'' states have smooth phase throughout the BZ (see Fig.~\ref{fig-pf-smooth}).
In addition, because $\epsilon_\uparrow({\bf k}) = \epsilon_\downarrow({\bf k})$ at each ${\bf k}$, the rotated states $\{ |\ell 1 {\bf k} \rangle, |\ell 2 {\bf k} \rangle \}$ continue to be eigenstates of the Bloch Hamiltonian.

The rotated states lead to exponentially localized Wannier functions.
The price for evading the topological obstruction is that the resulting Wannier functions are no longer Kramers pairs
\begin{equation}
W_{ \ell 1, \alpha \uparrow}({\bf r}) \neq W_{ \ell 2, \alpha \downarrow}^*({\bf r}).
\end{equation}
However, since we are after exact bounds on spectral weight, localization is more important.
A localized basis means that we can use the Peierls phase presented in sec.~\ref{app-vecPotLatModel}. Besides, the projected interactions are guaranteed to be short-ranged.

We now compute the renormalized Hubbard interaction
\begin{equation}
\widetilde{U} = |U| \sum_{ {\bf i}^\prime \alpha} \left( |W_{\ell 1,\alpha \uparrow}({\bf i}^\prime)|^2 |W_{\ell 2, \alpha \downarrow}({\bf i}^\prime)|^2 + |W_{\ell 1,\alpha \downarrow}({\bf i}^\prime)|^2 |W_{\ell 2, \alpha \uparrow}({\bf i}^\prime)|^2 - 2 \text{Re}\left[ W^*_{\ell 1,\alpha \uparrow}({\bf i}^\prime) W_{\ell 1,\alpha \downarrow}({\bf i}^\prime) W_{\ell 2,\alpha \uparrow}({\bf i}^\prime) W^*_{\ell 2,\alpha \downarrow}({\bf i}^\prime) \right]\right). \label{pf-utilde}
\end{equation}
For this model, we find $\widetilde{U} = 0.087 |U|$. We choose to work with $|U|$ such that $w \ll \widetilde{|U|} \ll E_0$.
\rev{Following the same strategy as before, we obtain the low-energy Hamiltonian
\begin{equation}
 \mathcal{H}_{\text{low}} = -|\widetilde{U}| \sum\limits_{ {\bf i} } c^{\dag}_{ {\bf i} \ell 1 } c^{\dag}_{ {\bf i} \ell 2 } c^{\phantom\dag}_{ {\bf i} \ell 2 } c^{\phantom\dag}_{ {\bf i} \ell 1 }    -  \sum\limits_{ {\bf i}, {\bf j} } \mathcal{K}_{ij} c^{\dag}_{ {\bf i} \ell 1 } c^{\dag}_{ {\bf i} \ell 2 } c^{\phantom\dag}_{ {\bf j} \ell 2 } c^{\phantom\dag}_{ {\bf j} \ell 1 } + \mathcal{H}_{\rho\rho} + \mathcal{H}_{\uparrow\downarrow}. \label{pf-low-Ham}
\end{equation}
where all interactions have a rather complicated structure. 
We focus here only on the pair hopping term that couples to ${\bf A}$, as explained before, and  
impacts the calculation of ${{\partial^2 \mathcal{H}_{\rm low}}/{\partial A_\mu^2}}$.
The pair hopping amplitude is given by}
\begin{eqnarray}
\mathcal{K}_{ij} &=& |U| \Big( \sum_{{\bf i}^\prime \alpha} W_{\ell 1,\alpha \uparrow}({\bf i}-{\bf i}^\prime) W_{\ell 2, \alpha \downarrow}({\bf i}-{\bf i}^\prime) W^*_{\ell 2, \alpha \downarrow}({\bf j}-{\bf i}^\prime)  W^*_{\ell 1, \alpha \uparrow}({\bf j}-{\bf i}^\prime) - \nonumber \\
&\phantom{=}& \phantom{\sum\limits_\alpha} W_{\ell 1,\alpha \uparrow}({\bf i}-{\bf i}^\prime) W_{\ell 2, \alpha \downarrow}({\bf i}-{\bf i}^\prime) W^*_{\ell 1, \alpha \downarrow}({\bf j}-{\bf i}^\prime)  W^*_{\ell 2, \alpha \uparrow}({\bf j}-{\bf i}^\prime) + ( \ell 1 \leftrightarrow \ell 2 ) \Big). \label{pf-kij-def}
\end{eqnarray}
It is worth noting that if $W_{\ell 1, \alpha\uparrow}({\bf i}) = W_{\ell 2, \alpha \downarrow}({\bf i})^*$ and $W_{ \ell 1, \alpha \downarrow}({\bf r}) = W_{ \ell 2, \alpha \uparrow}({\bf r}) = 0$ were true, Eqns.~(\ref{pf-utilde}),(\ref{pf-kij-def}) would collapse to the expressions we encountered for trivial bands.
We emphasize that this is precisely the topological obstruction arising from the non-trivial $\mathbb{Z}_2$ topology.

Despite the complicated structure, we find $\mathcal{K}_{ij}$ to be real, positive and short-ranged.
We can therefore continue with Cauchy-Schwarz inequality on the four-fermion expectation value and a triangle inequality, just as we did in Eqns.~(\ref{proj-D-1}-\ref{n-amgm}), to get
\begin{equation}
\widetilde{D}_\text{low} \leq \dfrac{\widetilde{n}}{4 N_c} \sum\limits_{ {\bf i}, {\bf j} } ( {\bf i} - {\bf j} )^2 |\mathcal{K}_{ij}| = 0.019 |U| \widetilde{n}.
\end{equation}
The bound is shown in Fig. 2 of main text.

\section{Mean-field theory}\label{app-mft}
We have shown comparisons of our bound to superfluid stiffness $D_s$ calculated within mean-field theory.
In this section, we will review the theory for completeness.
We will first address the gauge choice for multi-band Bloch Hamiltonian and derive Peierls phase in ${\bf k}$ space. That will allow us to find the current operators.
Then, we will decouple the interaction in the pairing channel and discuss the self-consistent gap and number equations, followed by
deriving an expression for $D_s$ using Kubo formula.
Finally, we will discuss why we can ignore Hartree-Fock corrections.

\subsection{Gauge for the Multi-band Bloch Hamiltonian}\label{app-gaugeBloch}
In the discussion so far, we have used the gauge, labeled by $\text{I}$, that satisfies $t^\text{I}_{\alpha\beta}({\bf k} + {\bf G}) = t^\text{I}_{\alpha\beta}({\bf k})$. This is different from the gauge typically used in ${\bf k}$-space calculations, where
\begin{equation}
t^\text{II}_{\alpha\beta}({\bf k} + {\bf G}) = V^\dag({\bf G}) t^\text{II}_{\alpha\beta}({\bf k}) V({\bf G}).
\end{equation}
The diagonal unitary matrix $V({\bf G})$ encodes the locations of the orbitals with entries $V({\bf G})_{\alpha\alpha} = e^{i {\bf G}\cdot \boldsymbol\tau_\alpha}$, where $\alpha = \{ 1, 2, \cdots M \}$ is the orbital label and $M$ is the number of orbitals (including spin).
The origin of this gauge can be traced back to the Fourier transform convention
\begin{equation}
d^{\phantom\dag}_{ {\bf k} \alpha} = \dfrac{1}{\sqrt{N_c}} \sum\limits_{ {\bf i} } e^{ i {\bf k}\cdot {\bf r}_{ i\alpha} } d^{\phantom\dag}_{ {\bf i} \alpha} \label{def-FT-2}
\end{equation}
where ${\bf r}_{i\alpha} = {\bf i} + \boldsymbol\tau_\alpha$. The $\alpha$ in the exponential makes it distinct from Eq.~\eqref{def-FT}. 
It further leads to the Bloch Hamiltonian
\begin{equation}
t^{\text{II}}_{\alpha\beta}({\bf k}) = \sum\limits_{ {\bf i}-{\bf j} } e^{ i {\bf k}\cdot( {\bf r}_{i\alpha} - {\bf r}_{j\beta}) } t_{\alpha\beta}( |{\bf r}_{i\alpha} - {\bf r}_{j\beta}|).
\end{equation}
that has the convenient property that under Peierls phase substitution (see Eq.~\eqref{vp1}), it transforms as $t^{\text{II}}_{\alpha\beta}({\bf k}) \rightarrow t^{\text{II}}_{\alpha\beta}({\bf k}-e{\bf A})$. We emphasize that this is not the case for gauge-I.
The \emph{minimal} substitution allows one to write current operators (which are derivatives with ${\bf A}$) in terms of derivatives with the crystal momentum ${\bf k}$. 
We have used gauge-II extensively in our earlier work \cite{HazraPRX2019}, which lead to ~(1) of the main text.
We will work in gauge-II to set up the mean-field Hamiltonian as well.

\subsection{Superfluid Stiffness}\label{app-mean-field-Ds}
The attractive Hubbard interaction is given by
\begin{equation}
    \mathcal{H}_\text{int} = - |U| \sum\limits_{ {\bf i} \alpha } \hat{n}_{ {\bf i}\alpha \uparrow } \hat{n}_{ {\bf i}\alpha \downarrow } \label{full-Int-negu}
\end{equation}
where $\hat{n}_{ {\bf i}\alpha \sigma } = d^\dag_{ {\bf i} \alpha\sigma} d^{\phantom\dag}_{ {\bf i} \alpha\sigma}$ is the density operator.
We decouple the interactions in the Cooper channel $-|U|\hat{n}_{ {\bf i} \alpha \uparrow } \hat{n}_{ {\bf i} \alpha \downarrow } \approx \Delta_{{\bf i} \alpha} d^\dag_{ {\bf i} \alpha \uparrow } d^\dag_{ {\bf i} \alpha \downarrow } + \text{h.c.} - |\Delta_{{\bf i} \alpha}|^2/|U|$ and find the mean-field Hamiltonian
\begin{equation}
\mathcal{H}_{\text{MF}} = \sum\limits_{ {\bf k} } \Phi^\dag_{ {\bf k} } \mathcal{H}^{ {\rm BdG }}({\bf k}) \Phi^{\phantom\dag}_{ {\bf k} } = \sum\limits_{ {\bf k} } \Phi^\dag_{ {\bf k} } \begin{pmatrix}
t_{\uparrow }({\bf k}) - \mu & \Delta \\ \Delta^\dag & \mu - t^T_{\downarrow }(-{\bf k})
\end{pmatrix}  \Phi^{\phantom\dag}_{ {\bf k} }  \label{ham-MF}
\end{equation}
where $\Delta$ in Eq.~\eqref{ham-MF} is an $M\times M$ matrix given by $\text{diag}[\Delta_\alpha, \cdots]$ and $\Phi^{\phantom\dag}_{ {\bf k} } = ( d^{\phantom\dag}_{ {\bf k} A \uparrow}, \cdots, d^{\dag}_{ -{\bf k} A \downarrow}, \cdots )^T$. 
We have restricted ourselves to the spatially uniform ansatz $\Delta_\alpha = - |U| \left\langle \sum\limits_{ {\bf i} } d^{\phantom\dag}_{ {\bf i}\alpha\downarrow} d^{\phantom\dag}_{ {\bf i}\alpha\uparrow} \right\rangle$. These gaps will be found self-consistently
\begin{equation}
\Delta_\alpha = -|U| \sum\limits_{ a {\bf k} } f^0[ E_a({\bf k})] \left\langle a {\bf k} \left| \dfrac{\partial }{\partial \Delta_\alpha} \mathcal{H}^{ {\rm BdG }}({\bf k}) \right| a {\bf k} \right\rangle \label{gap-eqn}
\end{equation}
where $a$ is a Boguliubov band label and $\mathcal{H}^{ {\rm BdG }}_{\bf k} | a{\bf k} \rangle = E_a({\bf k}) | a{\bf k} \rangle$. $f^0[\epsilon]$ is the Fermi function.
Along with the gaps, chemical potential $\mu$ should also be made to satisfy the number equation
\begin{equation}
n =  \sum\limits_{ a {\bf k} } f^0[ E_a({\bf k})]  \label{number-eqn}
\end{equation}
where $n$ is the total density of electrons.
Overall, there are $M+1$ equations that need to be solved self-consistently.

Since $D_s$ is a transverse current-current response function, we need to find the current operators. As the interactions are on-site, an external vector potential couples only to the kinetic part
\begin{equation}
\mathcal{H}^{ {\rm BdG }}({\bf k}) \longrightarrow \begin{pmatrix}
t_\uparrow({\bf k}- e{\bf A}) -\mu & \Delta \\ \Delta^\dag & \mu - t^T(-{\bf k}-e{\bf A})
\end{pmatrix}.
\end{equation}
We simplify the BdG matrix by invoking time reversal symmetry $t_{\uparrow }({\bf k}) = t^T_{\downarrow }(-{\bf k}) = t({\bf k})$ to write
\begin{equation}
\mathcal{H}^{ {\rm BdG }}({\bf k}) \longrightarrow \begin{pmatrix}
t({\bf k}- e{\bf A})-\mu & \Delta \\ \Delta^\dag & \mu - t({\bf k}+e{\bf A})
\end{pmatrix}
\end{equation}
that gives the current operators
\begin{eqnarray}
{\bf j}^P_\mu &=& e \sum\limits_{ {\bf k} } \Phi^\dag_{ {\bf k} } \left(\gamma^z \partial_\mu\mathcal{H}^{ {\rm BdG }}({\bf k}) \right)\Phi^{\phantom\dag}_{ {\bf k} } \\
{\bf j}^D_{\mu} &=& e^2 \sum\limits_{ {\bf k} } \Phi^\dag_{ {\bf k} } \left( \partial_{\mu\nu} \mathcal{H}^{ {\rm BdG }}({\bf k}) \right)\Phi^{\phantom\dag}_{ {\bf k} } {\bf A}_\nu.
\end{eqnarray}

\begin{table}
\centering
\begin{tabular}{| c | c | c | c | c | c | c |}
\hline
&&&&&& \\
& Band parameters & Spread functional & Minimum spread  & Mean-field  & QMC  & Exact bound \\
& & $\Omega$ & $\Omega_I = \text{Tr} \sum\limits_{\bf k}g({\bf k})$ & $D_s/|U|$ & $D_s/|U|$ &  $\widetilde{D}_\text{low}/|U|$\\
&&&&&& \\
\hline
&&&&&& \\
Lieb Lattice & $w=0$ & $0.12$ & $0.11$ & 0.021 & - & 0.03 \\
&$E_0 = t$&&&&& \\
&&&&&& \\
\hline
&&&&&& \\
Pi-Flux & $w=0.036t_1$ & $1.18$ & $1.03$ & 0.02 & $0.02 \pm 0.002$ \cite{HofmannPRB2020} & 0.019 \\
&$E_0 = 3.94 t_1$&&&&& \\
&&&&&& \\
\hline
\end{tabular}
\caption{Comparison of multi-band mean-field and QMC~\cite{HofmannPRB2020} estimates of $D_s$ for (a) Lieb lattice and (b) $\pi$-flux models. }
\end{table}

Here $\mu,\nu$ are spatial indices, $\partial_\mu = \partial/\partial k_\mu$ and $\gamma^z = \begin{pmatrix}
1 & 0\\ 0 & -1
\end{pmatrix} \otimes \mathbb{I}_{ M \times M}$ takes care of the opposite charges of particles and holes.
Next we use Kubo formula to calculate the tensor
\begin{equation}
[{\bf D_s}]_{\mu\nu} = [\widetilde{{\bf D}}]_{\mu\nu} - \dfrac{\hbar^2}{4e^2} [\chi_{JJ}^\text{MF}({\bf q}_\perp\rightarrow 0)]_{\mu\nu}
\end{equation}
where the diamagnetic part is
\begin{equation}
    [\widetilde{{\bf D}}]_{\mu\nu} = \dfrac{1}{4} \sum\limits_{ {\bf k}, a } f^0[E_a({\bf k})] \left\langle a {\bf k} | \partial_{\mu\nu} \mathcal{H}^{ {\rm BdG }}_{\bf k} | a {\bf k} \right\rangle \label{mf-Lieb-Dt}.
\end{equation}
and the paramagnetic current-current susceptibility is
\begin{equation}
    [\chi_{JJ}^\text{MF}]_{\mu\nu} = \sum\limits_{ {\bf k}, ab } \dfrac{f^0[E_a({\bf k})] - f^0[E_b({\bf k})-\mu] }{E_b({\bf k})- E_a({\bf k})} \langle a {\bf k} | \partial_\mu\mathcal{H}^{\text{BdG}}_{ {\bf k} } \gamma^z | b  {\bf k} \rangle \langle b  {\bf k} | \gamma^z \partial_\nu\mathcal{H}^{\text{BdG}}_{ {\bf k} } | a  {\bf k} \rangle \label{mf-Lieb-chi}.
\end{equation}
The pre-factor should be interpreted as $-\dfrac{\partial f}{\partial E} \Big|_{ E = E_a( {\bf k} ) }$ when $a=b$. 

For anisotropic systems in 2D, the critical temperature is controlled by the determinant of superfluid stiffness tensor (see appendix H of ref \cite{HazraPRX2019}). Therefore, in all the figures, we show
\begin{eqnarray}
D_s = \sqrt{ \text{det}[ {\bf D_s} ] }, \quad \widetilde{D} = \dfrac{1}{2} \text{Tr}[ \widetilde{{\bf D}} ].
\end{eqnarray}
Note that these definitions are compatible with the bound defined in Eq.~\eqref{def-dtilde-trace}:
\begin{equation}
D_s = \sqrt{\text{det}[{\bf D_s}]} \leq \sqrt{\text{det}[\widetilde{{\bf D}}_\text{low}]} \leq \dfrac{1}{2} \text{Tr}[ \widetilde{{\bf D}}_\text{low}]  = \widetilde{ D}_\text{low}.
\end{equation}

In our numerical scheme, we first find $D_s(\Delta, \mu)$ and then use the solutions of Eq.~\eqref{gap-eqn} and Eq.~\eqref{number-eqn} to calculate $D_s(|U|,n)$.

\rev{
\subsection{Hartree-Fock Corrections for Projected Hamiltonian}\label{app-HF-corr}

Finally, we turn to the question of Hartree-Fock (HF) corrections in the low-energy sector and whether that 
leads to additional spectral weight by giving rise to single-particle dispersion in the flat band.
Given the spin structure of the Hubbard Hamiltonian, there is no Fock correction. We show below that the Hartree term leads to 
at most a chemical potential shift for the models considered.  This does not impact our results, since we work at a {\it fixed density} with a 
partially filled flat band

We start with the full multi-band Hamiltonian of Eq.~\eqref{Hfull-3} and Eq.~\eqref{def-Vell} with the attractive Hubbard interaction.
The Hartree correction is then given by 
\begin{equation}
-|U|\sum\limits_{ {\bf k}, {\bf k}^\prime, {\bf q}, \ell, m } \left( \sum\limits_{ \alpha } U^{\phantom\dag}_{\ell \alpha }( {\bf k}-{\bf q}) U^\dag_{\alpha \ell }( {\bf k} )  U^{\phantom\dag}_{m \alpha }( {\bf k}^\prime+{\bf q} ) U^\dag_{ \alpha m}( {\bf k}^\prime) \right)  
c^{ \dag}_{ {\bf k}-{\bf q} \ell \uparrow} c^{ \phantom\dag}_{ {\bf k} \ell \uparrow} 
\left\langle c^{ \dag}_{ {\bf k}^\prime+{\bf q} m \downarrow } c^{ \phantom\dag}_{ {\bf k}^\prime m\downarrow} \right\rangle.
\end{equation}
Now $\left\langle c^{ \dag}_{ {\bf k}^\prime+{\bf q} m \downarrow } c^{ \phantom\dag}_{ {\bf k}^\prime m\downarrow} \right\rangle = \delta_{{\bf q},0}$
for a filled band $m$ and vanishes for an empty band. 

This correction {\it would appear to} lead to an interaction-induced dispersion 
\begin{equation}
\sum\limits_{ {\bf k},\sigma } \widetilde{\epsilon}_{\ell}({\bf k}) c^{ \dag}_{ {\bf k} \ell \sigma} c^{ \phantom\dag}_{ {\bf k} \ell \sigma}
\ \ \ \ {\rm with} \ \ \ \
\widetilde{\epsilon}_{\ell}({\bf k}) = -|U|\sum\limits_{ {\bf k}^\prime, \alpha, m} U^{\phantom\dag}_{\ell \alpha }( {\bf k}) U^\dag_{\alpha \ell }( {\bf k} )  U^{\phantom\dag}_{m \alpha }( {\bf k}^\prime ) U^\dag_{ \alpha m}( {\bf k}^\prime)  
\label{induced-disp}
\end{equation}
where the sum over $m$ only includes filled bands. This effect is absent in the $\pi$-flux model, where there are no filled bands below the flat bands. 
Also note that the effect is finite only when the filled band, $m$, and the active band, $\ell$, share support on a common orbital, $\alpha$.

For the Lieb lattice, we now show that although $\widetilde{\epsilon}_{\ell}({\bf k})$ is non-zero, it is just a ${\bf k}$-independent 
constant leading to a chemical potential shift.
From the Bloch Hamiltonian Eq.~\eqref{Lieb-full-K} we find the normalized single-particle wave functions.
The flat-band wave function Eq.~\eqref{lief-flat-band-waveF} can be written as
\begin{equation}
U_{\ell A}({\bf k}) = 0,\quad U_{\ell B}({\bf k}) = \dfrac{1}{\mathcal{N}({\bf k})}f^*(k_y), \quad U_{\ell C}({\bf k}) = -\dfrac{1}{\mathcal{N}({\bf k})} f(k_x)
\end{equation}
while that of the lower (fully occupied) band is given by
\begin{equation}
U_{m A}({\bf k}) = \dfrac{1}{\sqrt{2}},\quad U_{m B}({\bf k}) = \dfrac{1}{2\mathcal{N}({\bf k})} f^*(k_x), \quad U_{m C}({\bf k}) = \dfrac{1}{2\mathcal{N}({\bf k})} f(k_y).
\end{equation}
where $\mathcal{N}({\bf k})= \sqrt{|f(k_x)|^2 + |f(k_y)|^2}$ and $A,B,C$ are the sublattice labels. We can then explicitly check from
Eq.~\eqref{induced-disp} that 
$\widetilde{\epsilon}_{\ell}({\bf k})  = {-|U|}/{4}$ is just a constant.
We can gain further insight into the ${\bf k}$-independence of this result by expressing Eq.~\eqref{induced-disp} in real space using WFs.
The constant Hartree shift results from the flat-band having zero weight on the $A$ sublattice and 
 the probabilities of finding an electron on the $B$ or $C$ sublattice being the same by symmetry.
}

\end{document}